\documentclass[final, 3p]{elsarticle}
\usepackage{lineno}
\usepackage[export]{adjustbox}
\usepackage{float}
\usepackage{import}
\usepackage{subfig}
\usepackage{amsmath}
\usepackage{amssymb}
\usepackage{color}
\usepackage{bm}
\usepackage{tikz}
\usepackage{ifthen} 
\usepackage{soul} 

\usepackage[linesnumbered,ruled,lined,commentsnumbered]{algorithm2e}

\usepackage{etoolbox}
\newcommand*\linenomathpatch[1]{%
  \cspreto{#1}{\linenomath}%
  \cspreto{#1*}{\linenomath}%
  \cspreto{end#1}{\endlinenomath}%
  \cspreto{end#1*}{\endlinenomath}%
}

\linenomathpatch{equation}
\linenomathpatch{gather}
\linenomathpatch{multline}
\linenomathpatch{align}
\linenomathpatch{alignat}
\linenomathpatch{flalign}

\title{Scalable Parallel Linear Solver for Compact Banded Systems on Heterogeneous Architectures}

\author[1]{\small Hang Song\corref{cor1}}
\ead{songhang@stanford.edu}

\author[1]{\small Kristen V. Matsuno}
\ead{kmatsuno@stanford.edu}

\author[1]{\small Jacob R. West}
\ead{jrwest@stanford.edu}

\author[2]{\small Akshay Subramaniam}
\ead{akshays@stanford.edu}

\author[2]{\small Aditya S. Ghate}
\ead{aditya90@stanford.edu}

\author[1,2]{\small Sanjiva K. Lele}
\ead{lele@stanford.edu}

\cortext[cor1]{Corresponding author}

\address[1]{Department of Mechanical Engineering, Stanford University, Stanford, CA 94305, USA}
\address[2]{Department of Aeronautics \& Astronautics, Stanford University, Stanford, CA 94305, USA}

\begin{document}
\begin{abstract}
    A scalable algorithm for solving compact banded linear systems on distributed memory architectures is presented. The proposed method factorizes the original system into two levels of memory hierarchies, and solves it using parallel cyclic reduction on both distributed and shared memory. This method has a lower communication footprint across distributed memory partitions compared to conventional algorithms involving data transpose or re-partitioning. The algorithm developed in this work is generalized to cyclic compact banded systems with flexible data decompositions. For cyclic compact banded systems, the method is a direct solver with a deterministic operation and communication counts depending on the matrix size, its bandwidth, and the partition strategy. The implementation and runtime configuration details are discussed for performance optimization. Scalability is demonstrated on the linear solver as well as on a representative fluid mechanics application problem, in which the dominant computational cost is solving the cyclic tridiagonal linear systems of compact numerical schemes on a 3D periodic domain. The algorithm is particularly useful for solving the linear systems arising from the application of compact finite difference operators to a wide range of partial differential equation problems, such as but not limited to the numerical simulations of compressible turbulent flows, aeroacoustics, elastic-plastic wave propagation, and electromagnetics. It alleviates obstacles to their use on modern high performance computing hardware, where memory and computational power are distributed across nodes with multi-threaded processing units.
\end{abstract}
\maketitle

\section{Introduction}

In the past few decades, the use of graphics processing units (GPUs) in scientific computing has emerged as an attractive option to significantly accelerate various algorithms. The transition of several leadership class computing platforms to such heterogeneous architectures underscores the importance of numerical methods which can take full advantage of these nodes' parallel nature. The methods for solving certain linear systems presented in this work are well-suited for not only GPUs, but also platforms with hybrid memory management, and can take advantage of systems with distributed memory combined with multithreading.

In multiscale physics problems, such as simulations of compressible turbulent flow, the resolution of both large and small scales on a discrete grid is essential. Similarly, computational applications involving hydrodynamic instabilities and wave-propagation, such as in aeroacoustics, solid mechanics, and electromagnetics, require numerical discretizations with very low dispersion and dissipation errors. High order numerical methods have become increasingly attractive to tackle such problems since they provide high solution fidelity at a manageable computational cost \cite{colonius2004computational}. Differentiation using compact finite difference schemes and elliptic solves using spectral methods can be represented discretely as compact banded matrices, and are prime candidates for such multiscale computations due to their increased performance in the high wavenumber regime \cite{lele1992compact,gottlieb1977numerical}. The desirable performance of compact schemes for resolving large ranges of scales has been demonstrated in incompressible \cite{laizet2009high,simens2009high,ghate2017subfilter,uzun2018large} and compressible \cite{tritschler2014richtmyer,ryu2014turbulence,jagannathan2016reynolds,olson2011nonlinear} turbulent flows, aeroacoustics \cite{bodony2005using, wolf2012convective} as well as multiphysics applications with complex physical phenomena \cite{ghaisas2018unified, shang1999high}. These higher order finite differences are computed as a linear system with tridiagonal or other compact banded matrices. As derived by Lele \cite{lele1992compact}, the tridiagonal schemes for collocated first order derivatives, $f^\prime$, at gridpoint $i$ with spacing $h=x_i-x_{i-1}$ are formulated as
    \begin{equation}
        \alpha f_{i-1}' + f_i' + \alpha f_{i+1}' = b\frac{f_{i+2}-f_{i-2}}{4h} + a\frac{f_{i+1}-f_{i-1}}{2h}
    \end{equation}
Similarly, interpolation between values on collocated and staggered grids can also be formulated as a tridiagonal matrix, where $f$ is the original field and $f^I$ is the interpolated field \cite{nagarajan2003robust}:
    \begin{equation}
        \hat{\alpha} f^I_{i-1} + f^I_i + \hat{\alpha} f^I_{i+1} = b\frac{f_{i+3/2}+f_{i-3/2}}{2} + a\frac{f_{i+1/2}+f_{i-1/2}}{2}
    \end{equation}
For strong shock-turbulence interaction problems, the compact shock capturing schemes combined with Riemann solver have been proved to be both robust and less dissipative \cite{wong2017high,subramaniam2019high}. For such schemes, block tridiagonal (or wider banded) systems will be formed.

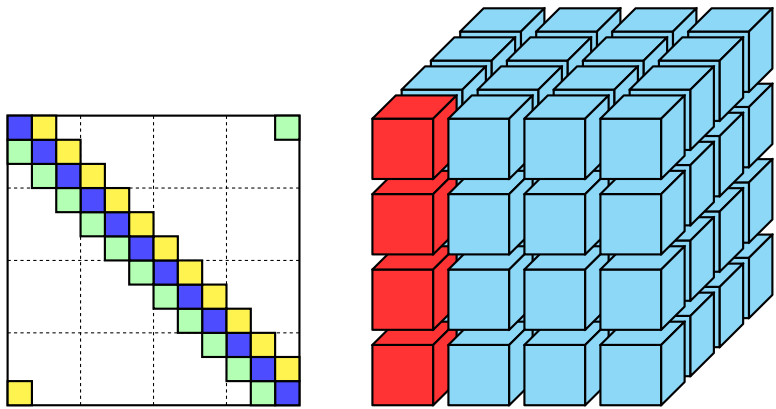
\begin{figure}[htbp]
    \centering
    \scalebox{0.4}{
\tikzset{
    thick/.style={line width=2.0pt},
    thin/.style={line width=1.0pt},
    regular block/.pic={
        \draw[fill=cyan!40!white, draw=black, thick, rounded corners=0.2pt] (0,0,0) -- (2,0, 0) -- (2,2, 0) -- (0,2, 0) -- cycle;
        \draw[fill=cyan!40!white, draw=black, thick, rounded corners=0.2pt] (0,2,0) -- (2,2, 0) -- (2,2,-2) -- (0,2,-2) -- cycle;
        \draw[fill=cyan!40!white, draw=black, thick, rounded corners=0.2pt] (2,0,0) -- (2,0,-2) -- (2,2,-2) -- (2,2, 0) -- cycle;
    },
    highlighted block/.pic={
        \draw[fill=red!80!white, draw=black, thick, rounded corners=0.2pt] (0,0,0) -- (2,0, 0) -- (2,2, 0) -- (0,2, 0) -- cycle;
        \draw[fill=red!80!white, draw=black, thick, rounded corners=0.2pt] (0,2,0) -- (2,2, 0) -- (2,2,-2) -- (0,2,-2) -- cycle;
        \draw[fill=red!80!white, draw=black, thick, rounded corners=0.2pt] (2,0,0) -- (2,0,-2) -- (2,2,-2) -- (2,2, 0) -- cycle;
    },
    grid partitioning/.pic={
        \foreach \i in {0,...,3} {
            \foreach \j in {0,...,3} {
                \foreach \k in {0,...,3} {
                    \ifthenelse{\i = 0 \AND \k = 3}
                        {\pic at (\i*2.5, \j*2.5, \k*2.5) {highlighted block};}
                        {\pic at (\i*2.5, \j*2.5, \k*2.5) {regular block};}
                }
            }
        }
        
        \draw[thick](-12, 0, 7.5) rectangle (-2.4, 9.6, 7.5);
        \foreach \i in {1,...,3} {
            \draw[thin, dashed] (\i*2.4-12, 0, 7.5) -- (\i*2.4-12, 9.6, 7.5);
            \draw[thin, dashed] (-12, \i*2.4, 7.5) -- (-2.4, \i*2.4, 7.5);
        }
        \foreach \i in {0,...,11} {
            \draw[thick, fill=blue!70!white] (-\i*0.8-2.4,    \i*0.8, 7.5) rectangle ++ (-0.8,0.8,0);
            \ifnum \i = 11
                \draw[thick, fill=green!30!white] (-\i*0.8-2.4+8.8, \i*0.8, 7.5) rectangle ++ (-0.8,0.8,0);
            \else
                \draw[thick, fill=green!30!white] (-\i*0.8-2.4-0.8, \i*0.8, 7.5) rectangle ++ (-0.8,0.8,0);
            \fi
            \ifnum \i = 0
                \draw[thick, fill=yellow!80!white] (-\i*0.8-2.4-8.8, \i*0.8, 7.5) rectangle ++ (-0.8,0.8,0);
            \else
                \draw[thick, fill=yellow!80!white] (-\i*0.8-2.4+0.8, \i*0.8, 7.5) rectangle ++ (-0.8,0.8,0);
            \fi
        }
    },
}

\begin{tikzpicture}
    \pic at (0,0,0) {grid partitioning};
\end{tikzpicture}}
    \caption{Structure of cyclic tridiagonal linear system (left) and 3D grid decomposition (right). Each highlighted chunk in the featured grid decomposition corresponds with the section of rows in the linear system shown.}
    \label{fig:grid-partition}
\end{figure}

Multiphysics solvers for structured, Eulerian grids in a multidimensional domain may be decomposed as shown in Figure \ref{fig:grid-partition}, with each processor given access to a single chunk of the global domain. This decomposition is particularly useful for fixed, structured, Cartesian grids since the grid chunks on each processor can easily be determined from the decomposition layout using simple algebra. This method of grid decomposition facilitates workload distribution, and works particularly well for architectures with a distributed memory layout. Operations such as derivatives or interpolation along one dimension involve communication across a single row or column of grid partitioning, such as the chunks highlighted in red in Figure \ref{fig:grid-partition}. As shown in dotted lines in the matrix, sections of the matrix are initially distributed among several processors or nodes; the linear solver of this system relies heavily on its communication requirements. This work presents a linear solver for compact banded systems with highly scalable properties. First, a brief review of cyclic reduction (CR) and parallel cyclic reduction (PCR) for banded matrices is given. Section \ref{sec:GeneralizedPCR} illustrates the generalized PCR for generic acyclic compact banded systems, which serves as a building block of the proposed algorithm. Section \ref{sec:linearSolver} describes in detail the solution process for tridiagonal matrices of arbitrary size on an arbitrary number of processors, followed by an analytical extension of the method for other compact banded matrices. Section \ref{sec:ImplementationDetails} provides additional implementation details to improve performance. In Section \ref{sec:performance}, a demonstration is provided of the computational performance of the linear solver and its use to solve the Navier-Stokes equations for the Taylor-Green vortex problem.

CR is a popular direct solve algorithm for structured matrix linear systems, particularly block tridiagonal linear systems \cite{gander1997cyclic}. It recursively reduces a linear system to half-size sub-systems until the size of the sub-system (typically $1\times1$) makes it affordable to solve. Once the sub-system is solved, the result can propagate backward to the parent system to solve for the remaining unknowns. Hockney \cite{hockney1965fast} initially derived CR in combination with the fast Fourier transform as an alternative algorithm for iterative solvers for the Poisson equation. Later, Buzbee et al. \cite{buzbee1970direct} presented a unified  formulation and generalization of Hockney's CR and Buneman's \cite{buneman1969compact} algorithm, which had mathematically equivalent reduction processes but differences in round-off errors and stability. Sweet \cite{sweet1974generalized,sweet1977cyclic} further generalized CR from matrices with block sizes of power-of-two to matrices of arbitrary block sizes. Similarly, Swarztrauber \cite{swarztrauber1974direct} also generalized CR for tridiagonal systems associated with separable elliptic equations. A parallel variant of CR, also known as PCR, was introduced by Hockney and Jesshope \cite{hockney1981parallel}. In the PCR process, the upper and lower off-diagonal elements of both the even and odd indexed rows of a tridiagonal matrix are  simultaneously eliminated by the previous and the next rows in one step of reduction. As a consequence, it splits a system into two half-size sub-systems in each step of PCR. The communication pattern of an $8\times8$ non-cyclic tridiagonal system is shown in Figure \ref{fig:pcr-tree}. After enough recursive splitting, all the sub-systems are of effectively trivial size, e.g. $1\times1$ in the bottom layer of Figure \ref{fig:pcr-tree}, to solve all the unknowns in parallel. This means that PCR solves the linear system in a single forward pass and does not require a backward substitution phase.

\begin{figure}[htbp]
    \centering
    \scalebox{0.4}{\tikzset{
    thick/.style={line width=2.5pt},
    thin/.style={line width=1.5pt},
}
\begin{tikzpicture}
    \foreach \i in {1,...,8} {
        \draw[thick, fill=green!20!white] (\i*2, 0) circle (0.7);
    }
    \foreach \i in {1,...,4} {
        \draw[thick, fill=  blue!40!white] (\i*4  ,-4) circle (0.7);
        \draw[thick, fill=yellow!50!white] (\i*4-2,-4) circle (0.7);
    }
    \foreach \i in {1,...,2} {
        \draw[thick, fill=purple!50!white] (\i*8  ,-8) circle (0.7);
        \draw[thick, fill=  cyan!10!white] (\i*8-2,-8) circle (0.7);
        \draw[thick, fill=  lime!60!white] (\i*8-4,-8) circle (0.7);
        \draw[thick, fill= brown!30!white] (\i*8-6,-8) circle (0.7);
    }
    \draw[thick, fill=  teal!50!white] ( 2,-12) circle (0.7);
    \draw[thick, fill= olive!50!white] ( 4,-12) circle (0.7);
    \draw[thick, fill=orange!70!white] ( 6,-12) circle (0.7);
    \draw[thick, fill=yellow!90!white] ( 8,-12) circle (0.7);
    \draw[thick, fill=purple!20!white] (10,-12) circle (0.7);
    \draw[thick, fill=  cyan!30!white] (10,-12) circle (0.7);
    \draw[thick, fill=violet!40!white] (12,-12) circle (0.7);
    \draw[thick, fill= green!50!white] (14,-12) circle (0.7);
    \draw[thick, fill=  blue!20!white] (16,-12) circle (0.7);
    \foreach \i in {1,...,8} {
        \foreach \j in {0,...,2} {
            \draw[->, >=stealth, thick] (\i*2, -4*\j-1) --++ (0,-2);
            \pgfmathparse{\i*2-2*2^\j>=2 ? 1 : 0}
            \ifnum \pgfmathresult>0
                \draw[->, >=stealth, thick] (\i*2, -4*\j-1) --++ (-2*2^\j,-2);
            \fi
            \pgfmathparse{\i*2+2*2^\j<=16 ? 1 : 0}
            \ifnum \pgfmathresult>0
                \draw[->, >=stealth, thick] (\i*2, -4*\j-1) --++ (2*2^\j,-2);
            \fi
        }
        \foreach \j in {0,...,3} {
            \node [anchor=center] at (\i*2, -4*\j) {\LARGE\i};
        }
    }
\end{tikzpicture}}
    \caption{Communication pattern of PCR for an $8\times8$ non-cyclic tridiagonal system. The sub-systems in each step are grouped by the same colors.}
    \label{fig:pcr-tree}
\end{figure}

Recent works have optimized both CR and PCR for modern parallel computer architectures, and have achieved considerable performance improvements for specific applications. For example, a GPU implementation is suggested by Zhang et al. \cite{zhang2010fast}, and the works of Hirshman et al. \cite{hirshman2010bcyclic} and Seal et al. \cite{seal2013revisiting} improve the algorithm for block tridiagonal systems with large dense blocks. Nevertheless, most of the general PCR solvers are implemented for shared memory data access, and few improved algorithms have comprehensively considered data partitioning for distributed memory. The parallel linear solver developed in this paper is based on the concept of PCR to solve the banded system, and optimized for the grid decomposition on the distributed memory shown in Figure \ref{fig:grid-partition}. These banded systems typically are (block) tridiagonal or (block) pentadiagonal systems, but the present algorithm can be extended to wider bandwidths.

\section{Generalized parallel cyclic reduction method}\label{sec:GeneralizedPCR}

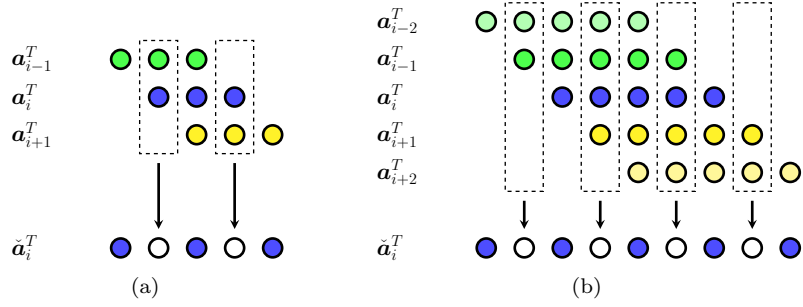
\begin{figure}[htbp]
    \centering
    \subfloat[]
    {\scalebox{0.5}{\tikzset{
    thick/.style={line width=2.0pt},
    thin/.style={line width=1.0pt},
    block prev2/.pic={
        \draw[fill=green!30!white, draw=black, thick] (0,0) circle (0.25);
    },
    block prev1/.pic={
        \draw[fill=green!70!white, draw=black, thick] (0,0) circle (0.25);
    },
    block central/.pic={
        \draw[fill=blue!70!white, draw=black, thick] (0,0) circle (0.25);
    },
    block next1/.pic={
        \draw[fill=yellow!90!white, draw=black, thick] (0,0) circle (0.25);
    },
    block next2/.pic={
        \draw[fill=yellow!50!white, draw=black, thick] (0,0) circle (0.25);
    },
    block empty/.pic={
        \draw[fill=white, draw=black, thick] (0,0) circle (0.25);
    },
    block diagonal/.pic={
        \draw[fill=red!80!white, draw=black, thick] (0,0) circle (0.25);
    },
    box3/.pic={
        \draw[dashed, thin] (-0.5,1.5) rectangle (0.5,-1.5);
    },
    box5/.pic={
        \draw[dashed, thin] (-0.5,2.5) rectangle (0.5,-2.5);
    },
}

\begin{tikzpicture}
    \foreach \i in {-1,...,1} {
        \pic at (\i-1, 1) {block prev1};
        \pic at (\i  , 0) {block central};
        \pic at (\i+1,-1) {block next1};
    }
    \foreach \i in {-1,...,1} {
        \pic at (2*\i, -4) {block central};
    }
    \foreach \i in {-1,1} {
        \pic at (\i, -4) {block empty};
        \pic at (\i,0) {box3};
        \draw[->, >=stealth, thick] (\i,-1.75) -- (\i,-3.5);
    }
    \node[anchor=west] at (-5, 0) {\LARGE ${\bm a}^T_{i}$};
    \node[anchor=west] at (-5,-1) {\LARGE ${\bm a}^T_{i+1}$};
    \node[anchor=west] at (-5, 1) {\LARGE ${\bm a}^T_{i-1}$};
    \node[anchor=west] at (-5,-4) {\LARGE $\check{\bm a}^T_{i}$};
\end{tikzpicture}}\label{fig:GPCR-tri}}
    \qquad\quad
    \subfloat[]
    {\scalebox{0.5}{\tikzset{
    thick/.style={line width=2.0pt},
    thin/.style={line width=1.0pt},
    block prev2/.pic={
        \draw[fill=green!30!white, draw=black, thick] (0,0) circle (0.25);
    },
    block prev1/.pic={
        \draw[fill=green!70!white, draw=black, thick] (0,0) circle (0.25);
    },
    block central/.pic={
        \draw[fill=blue!70!white, draw=black, thick] (0,0) circle (0.25);
    },
    block next1/.pic={
        \draw[fill=yellow!90!white, draw=black, thick] (0,0) circle (0.25);
    },
    block next2/.pic={
        \draw[fill=yellow!50!white, draw=black, thick] (0,0) circle (0.25);
    },
    block empty/.pic={
        \draw[fill=white, draw=black, thick] (0,0) circle (0.25);
    },
    block diagonal/.pic={
        \draw[fill=red!80!white, draw=black, thick] (0,0) circle (0.25);
    },
    box3/.pic={
        \draw[dashed, thin] (-0.5,1.5) rectangle (0.5,-1.5);
    },
    box5/.pic={
        \draw[dashed, thin] (-0.5,2.5) rectangle (0.5,-2.5);
    },
}

\begin{tikzpicture}
    \foreach \i in {-2,...,2} {
        \pic at (\i-2, 2) {block prev2};
        \pic at (\i-1, 1) {block prev1};
        \pic at (\i  , 0) {block central};
        \pic at (\i+1,-1) {block next1};
        \pic at (\i+2,-2) {block next2};
    }
    \foreach \i in {-2,...,2} {
        \pic at (2*\i, -4) {block central};
    }
    \foreach \i in {-3,-1,1,3} {
        \pic at (\i, -4) {block empty};
        \pic at (\i,0) {box5};
        \draw[->, >=stealth, thick] (\i,-2.75) -- (\i,-3.5);
    }
    \node[anchor=west] at (-7, 0) {\LARGE ${\bm a}^T_{i}$};
    \node[anchor=west] at (-7,-1) {\LARGE ${\bm a}^T_{i+1}$};
    \node[anchor=west] at (-7,-2) {\LARGE ${\bm a}^T_{i+2}$};
    \node[anchor=west] at (-7, 1) {\LARGE ${\bm a}^T_{i-1}$};
    \node[anchor=west] at (-7, 2) {\LARGE ${\bm a}^T_{i-2}$};
    \node[anchor=west] at (-7,-4) {\LARGE $\check{\bm a}^T_{i}$};
\end{tikzpicture}}\label{fig:GPCR-penta}}
    \caption{Example of one step in generalized PCR: \protect\subref{fig:GPCR-tri} tridiagonal system; \protect\subref{fig:GPCR-penta} penta-diagonal system. The colored circles are non-identically-zero entries and the uncolored circles are identically-zero entries.}
    \label{fig:generalized-pcr}
\end{figure}

Beyond the tridiagonal system, PCR can be easily generalized for a compact banded system with arbitrary bandwidth. In order to form two sub-systems grouped by the even and odd rows, each row in the parent system, after a reduction step, is staggered with a zero entry between any of the two non-zero entries on the diagonal and off-diagonals, as shown in Figure \ref{fig:generalized-pcr}. In the generalized PCR approach, the total number of neighbor rows involved to eliminate the entries in row $i$ equals the number of the off-diagonal elements. And the resulting row $i$ is the linear combination of row $i$ and the neighbor rows.

Let ${\bm a}_i^T$ be the $i$-th row vector in the parent matrix, and the reduction operation to obtain the $i$-th row vector in the resulted matrix, $\check{\bm a}_i^T$, can be expressed as
\begin{equation}
    \check{\bm a}_i^T = {\bm a}_i^T - \sum_{j=1}^{(w-1)/2}\left(k_{+j}{\bm a}_{i+j}^T + k_{-j}{\bm a}_{i-j}^T\right)
\end{equation}
where $w$ is the bandwidth of the compact banded system. During a reduction step, each of the zero staggered entries can be formed with a unique linear combination of the involved neighbor row vectors, as the boxed columns in Figure \ref{fig:generalized-pcr}. The coefficients, $k_{+j}$ and $k_{-j}$ can be solved from the linear system described in Equation (\ref{eqn:k_system_general}).
\begin{equation}
    \begin{bmatrix}
        \ddots & \ddots & \ddots & & & \\
        \cdots & a_{i-2,i-3} & a_{i-1,i-3} & \cdots & &\\
        \cdots & a_{i-2,i-1} & a_{i-1,i-1} & a_{i+1,i-1} & \cdots &\\
        & \cdots & a_{i-1,i+1} & a_{i+1,i+1} & a_{i+2,i+1} & \cdots \\
        & & \cdots & a_{i+1,i+3} & a_{i+2,i+3} & \cdots\\
        & & & \ddots & \ddots & \ddots
    \end{bmatrix}
    \begin{bmatrix}
        \vdots \\ k_{-2} \\ k_{-1} \\ k_{+1} \\ k_{+2} \\ \vdots
    \end{bmatrix}
    =
    \begin{bmatrix}
        \vdots \\ a_{i,i-3} \\ a_{i,i-1} \\ a_{i,i+1} \\ a_{i,i+3} \\ \vdots
    \end{bmatrix}
    \label{eqn:k_system_general}
\end{equation}
Specifically, for a tridiagonal parent system ($w=3$), $k_{+j}$ and $k_{-j}$ for each row $i$ are governed by a $2\times2$ diagonal system shown in Equation (\ref{eqn:k_system_tridiagonal}). For a penta-diagonal parent system ($w=5$), $a_{i,i-3} = a_{i,i+3} = 0$ for each row $i$, and $k_{+j}$ and $k_{-j}$ are governed by a $4\times4$ tridiagonal system described in Equation (\ref{eqn:k_system_penta-diagonal}).
\begin{equation}
    \begin{bmatrix}
        a_{i-1,i-1} & \\
        & a_{i+1,i+1}
    \end{bmatrix}
    \begin{bmatrix}
        k_{-1} \\ k_{+1}
    \end{bmatrix}
    =
    \begin{bmatrix}
        a_{i,i-1} \\ a_{i,i+1}
    \end{bmatrix}
    \label{eqn:k_system_tridiagonal}
\end{equation}

\begin{equation}
    \begin{bmatrix}
        a_{i-2,i-3} & a_{i-1,i-3} & & \\
        a_{i-2,i-1} & a_{i-1,i-1} & a_{i+1,i-1} & \\
        & a_{i-1,i+1} & a_{i+1,i+1} & a_{i+2,i+1} \\
        & & a_{i+1,i+3} & a_{i+2,i+3}
    \end{bmatrix}
    \begin{bmatrix}
        k_{-2} \\ k_{-1} \\ k_{+1} \\ k_{+2}
    \end{bmatrix}
    =
    \begin{bmatrix}
        0 \\ a_{i,i-1} \\ a_{i,i+1} \\ 0
    \end{bmatrix}
    \label{eqn:k_system_penta-diagonal}
\end{equation}

\section{Parallel linear solver for compact banded system}
\label{sec:linearSolver}
\begin{figure}[htbp]
    \centering
    \scalebox{0.4}{
\tikzset{
    thick/.style={line width=2.5pt},
    thin/.style={line width=1.5pt},
    middle block small/.pic={
        \draw[fill=magenta!50!white, draw=black, thick] (0,0) rectangle (1,-1);
    },
    left block small/.pic={
        \fill[magenta!50!white] (0,0) -- (1,-1) |- cycle;
        \draw[draw=black, thick] (-2,0) rectangle (1,-1);
    },
    right block small/.pic={
        \fill[magenta!50!white] (0,0) -- (1,-1) -| cycle;
        \draw[draw=black, thick] (0,0) rectangle (3,-1);
    },
    middle block large/.pic={
        \fill[magenta!50!white] (0,0) -- (1,0) -- (3,-2) -- (3,-3) -- (2,-3) -- (0,-1) -- cycle;
        \draw[draw=black, thick] (0,0) rectangle (3,-3);
    },
    left block large/.pic={
        \fill[magenta!50!white] (0,0) -- (1,-1) |- cycle;
        \draw[draw=black, thick] (0,0) rectangle (1,-3);
    },
    right block large/.pic={
        \fill[magenta!50!white] (0,-2) -- (1,-3) -| cycle;
        \draw[draw=black, thick] (0,0) rectangle (1,-3);
    },
    matrix partition middle/.pic={
         \pic at ( 0, 0) {middle block small};
         \pic at ( 1, 0) {right block small};
         \pic at ( 1,-1) {middle block large};
         \pic at ( 0,-1) {left block large};
    },
    matrix A/.pic={
        \foreach \i in {-1,...,1} {\pic at (\i*4,-\i*4) {matrix partition middle};}
        \pic at (-1, 0) {left block small};
        \pic at ( 3,-4) {left block small};
        \pic at ( 4,-1) {right block large};
        \pic at ( 0, 3) {right block large};
        \foreach \i in {-2,...,1} {\draw[thick, dashed] (-6, \i*4) -- (10, \i*4);}
        \foreach \i in {-1,...,2} {\draw[thick, dashed] (\i*4,-9) -- (\i*4, 5);}
        
        \node[anchor=center] at ( 2.5, -2.5) {\huge$\bm D_{i}$};
        \node[anchor=center] at ( 6.5, -6.5) {\huge$\bm D_{i+1}$};
        \node[anchor=center] at (-1.5,  1.5) {\huge$\bm D_{i-1}$};
        
        \node[anchor=center] at (-1.5, -3.3) {\huge$\bm L_{i}$};
        \node[anchor=center] at ( 2.5, -7.3) {\huge$\bm L_{i+1}$};
        \node[anchor=center] at (-5.5,  0.7) {\huge$\bm L_{i-1}$};
        \draw[->, >=stealth, thin, draw=blue] (-1.0, -3.0) to [out=60, in=-120] ( 0.5,-1.8);
        \draw[->, >=stealth, thin, draw=blue] ( 3.0, -7.0) to [out=60, in=-120] ( 4.5,-5.8);
        \draw[->, >=stealth, thin, draw=blue] (-5.0,  1.0) to [out=60, in=-120] (-3.5, 2.2);
        
        \node[anchor=center] at (6.5, -2.5) {\huge$\bm U_{i}$};
        \node[anchor=center] at (2.5,  1.5) {\huge$\bm U_{i-1}$};
        \draw[->, >=stealth, thin, draw=blue] (6.0, -2.8) to [out=-150, in=30] ( 4.5,-3.2);
        \draw[->, >=stealth, thin, draw=blue] (2.0,  1.2) to [out=-150, in=30] ( 0.5, 0.8);
        
        \node[anchor=center] at (-3.0, -1.8) {\huge$\widetilde{\bm L}_{i}$};
        \node[anchor=center] at ( 1.0, -5.8) {\huge$\widetilde{\bm L}_{i+1}$};
        \draw[->, >=stealth, thin, draw=blue] (-2.5, -1.5) to [out=10, in=-170] (-1.0, -0.5);
        \draw[->, >=stealth, thin, draw=blue] ( 1.5, -5.5) to [out=10, in=-170] ( 3.0, -4.5);
        
        \node[anchor=west] at ( 5.3, -0.8) {\huge$\widetilde{\bm U}_{i}$};
        \node[anchor=west] at ( 1.3,  3.2) {\huge$\widetilde{\bm U}_{i-1}$};
        \node[anchor=west] at ( 9.3, -4.8) {\huge$\widetilde{\bm U}_{i+1}$};
        \draw[->, >=stealth, thin, draw=blue] (5.3, -0.8) to [out=-180, in=10] ( 2.5,-0.5);
        \draw[->, >=stealth, thin, draw=blue] (1.3,  3.2) to [out=-180, in=10] (-1.5, 3.5);
        \draw[->, >=stealth, thin, draw=blue] (9.3, -4.8) to [out=-180, in=10] ( 6.5,-4.5);
        
        \node[anchor=center] at (-1.0, -1.8) {\huge$\widetilde{\bm D}_{i}$};
        \node[anchor=center] at ( 3.0, -5.8) {\huge$\widetilde{\bm D}_{i+1}$};
        \node[anchor=center] at (-5.0,  2.2) {\huge$\widetilde{\bm D}_{i-1}$};
        \draw[->, >=stealth, thin, draw=blue] (-0.5, -1.5) to [out=30, in=-150] ( 0.5,-0.5);
        \draw[->, >=stealth, thin, draw=blue] (-4.5,  2.5) to [out=30, in=-150] (-3.5, 3.5);
        \draw[->, >=stealth, thin, draw=blue] ( 3.5, -5.5) to [out=30, in=-150] ( 4.5,-4.5);
    },
    vector x block/.pic={
        \fill[green!30!white] (0, 0) rectangle (1, -4);
        \draw[thick] ( 0, 0) rectangle ( 1, -1);
        \draw[thick] ( 0,-1) rectangle ( 1, -4);
    },
    vector x/.pic={
         \pic at ( 0, 0) {vector x block};
         \pic at ( 0,-4) {vector x block};
         \pic at ( 0, 4) {vector x block};
         \draw[thick, dashed] (-1, 4) -- (2, 4);
         \draw[thick, dashed] (-1, 0) -- (2, 0);
         \draw[thick, dashed] (-1,-4) -- (2,-4);
         \draw[thick, dashed] (-1,-8) -- (2,-8);
         
         \node[anchor=west] at (2.0,  3.5) {\huge$\widetilde{\bm x}_{i-1}$};
         \node[anchor=west] at (2.0, -0.5) {\huge$\widetilde{\bm x}_{i}$};
         \node[anchor=west] at (2.0, -4.5) {\huge$\widetilde{\bm x}_{i+1}$};
         \node[anchor=west] at (2.0,  1.5) {\huge${\bm x}_{i-1}$};
         \node[anchor=west] at (2.0, -2.5) {\huge${\bm x}_{i}$};
         \node[anchor=west] at (2.0, -6.5) {\huge${\bm x}_{i+1}$};
         
         \draw[->, >=stealth, thin, draw=blue] (1.8, 3.5) to [out=-150, in= 30] (0.5, 3.5);
         \draw[->, >=stealth, thin, draw=blue] (1.8,-0.5) to [out=-150, in= 30] (0.5,-0.5);
         \draw[->, >=stealth, thin, draw=blue] (1.8,-4.5) to [out=-150, in= 30] (0.5,-4.5);
         
         \draw[->, >=stealth, thin, draw=blue] (1.8, 1.5) to [out= 150, in=-30] (0.5, 1.5);
         \draw[->, >=stealth, thin, draw=blue] (1.8,-2.5) to [out= 150, in=-30] (0.5,-2.5);
         \draw[->, >=stealth, thin, draw=blue] (1.8,-6.5) to [out= 150, in=-30] (0.5,-6.5);
    },
    vector b block/.pic={
        \fill[cyan!50!white] (0, 0) rectangle (1, -4);
        \draw[thick] ( 0, 0) rectangle ( 1, -1);
        \draw[thick] ( 0,-1) rectangle ( 1, -4);
    },
    vector b/.pic={
         \pic at ( 0, 0) {vector b block};
         \pic at ( 0,-4) {vector b block};
         \pic at ( 0, 4) {vector b block};
         \draw[thick, dashed] (-1, 4) -- (2, 4);
         \draw[thick, dashed] (-1, 0) -- (2, 0);
         \draw[thick, dashed] (-1,-4) -- (2,-4);
         \draw[thick, dashed] (-1,-8) -- (2,-8);
         
         \node[anchor=west] at (2.0,  3.5) {\huge$\widetilde{\bm b}_{i-1}$};
         \node[anchor=west] at (2.0, -0.5) {\huge$\widetilde{\bm b}_{i}$};
         \node[anchor=west] at (2.0, -4.5) {\huge$\widetilde{\bm b}_{i+1}$};
         \node[anchor=west] at (2.0,  1.5) {\huge${\bm b}_{i-1}$};
         \node[anchor=west] at (2.0, -2.5) {\huge${\bm b}_{i}$};
         \node[anchor=west] at (2.0, -6.5) {\huge${\bm b}_{i+1}$};
         
         \draw[->, >=stealth, thin, draw=blue] (1.8, 3.5) to [out=-150, in= 30] (0.5, 3.5);
         \draw[->, >=stealth, thin, draw=blue] (1.8,-0.5) to [out=-150, in= 30] (0.5,-0.5);
         \draw[->, >=stealth, thin, draw=blue] (1.8,-4.5) to [out=-150, in= 30] (0.5,-4.5);
         
         \draw[->, >=stealth, thin, draw=blue] (1.8, 1.5) to [out= 150, in=-30] (0.5, 1.5);
         \draw[->, >=stealth, thin, draw=blue] (1.8,-2.5) to [out= 150, in=-30] (0.5,-2.5);
         \draw[->, >=stealth, thin, draw=blue] (1.8,-6.5) to [out= 150, in=-30] (0.5,-6.5);
    },
}

\begin{tikzpicture}
    \pic at ( 0,0)   {matrix A};
    \pic at (13,0)   {vector x};
    \pic at (20,0)   {vector b};
    \node at ( 2,-10) {\Huge$\bm A$};
    \node at (13.5,-10) {\Huge$\bm x$};
    \node at (20.5,-10) {\Huge$\bm b$};
\end{tikzpicture}}
    \caption{Partitioned linear system.}
    \label{fig:partitioned_system}
\end{figure}
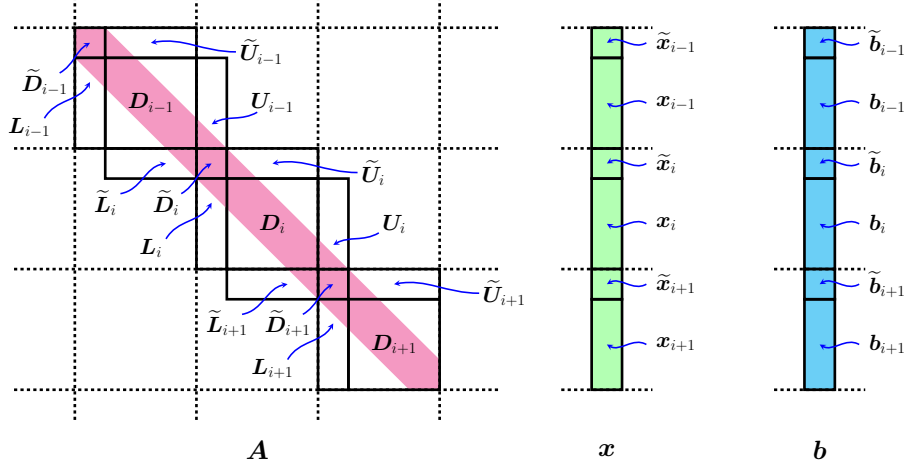

This section will introduce the parallel direct solver used for solving compact banded linear systems with the data partition on the distributed memory. Consistent with the grid decomposition pattern in Figure \ref{fig:grid-partition}, the compact banded linear system, ${\bm A}{\bm x} = {\bm b}$, is also correspondingly decomposed into a sparse block tridiagonal system \cite{subramaniam2018simulations} shown in Figure \ref{fig:partitioned_system}. The data in ${\bm x}$ and ${\bm b}$ are stored in the distributed memory. The subscripts in Figure \ref{fig:partitioned_system} indicate the rank of the aligned grid decomposition. Each rank has access to the data stored in its shared memory, the boundaries of which are indicated by dotted lines. $\widetilde{\bm D}_i$ is an $r\times r$ dense square matrix, whose dimension, $r$, is equal to half the number of off-diagonal bands in the linear system, $(w-1)/2$. For a tridiagonal system ($w=3$), $\widetilde{\bm D}_i$ is $1\times1$, and for a penta-diagonal system ($w=5$), $\widetilde{\bm D}_i$ is $2\times2$, etc. $\widetilde{\bm L}_i$ and $\widetilde{\bm U}_i$ are short, fat blocks, and ${\bm L}_i$ and ${\bm U}_i$ are tall, skinny blocks. ${\bm D}_i$ is a large, square, non-cyclic, banded block.

According to this grouping strategy, two equations are formed within each partition.
\begin{align}
    \widetilde{\bm L}_i {\bm x}_{i-1} + \widetilde{\bm D}_i\widetilde{\bm x}_i + \widetilde{\bm U}_i {\bm x}_i &= \widetilde{\bm b}_i
    \label{eqn:system1}
    \\
    {\bm L}_i \widetilde{\bm x}_i + {\bm D}_i{\bm x}_i + {\bm U}_i \widetilde{\bm x}_{i+1} &= {\bm b}_i
    \label{eqn:system2}
\end{align}
Assuming ${\bm D}_i$ is invertible -- which is true for the linear systems formed from compact schemes -- then ${\bm x}_i$ can be obtained if both $\widetilde{\bm x}_i$ and $\widetilde{\bm x}_{i-1}$ are known.
\begin{equation}
    {\bm x}_i = {\bm D}_i^{-1}\left[{\bm b}_i - {\bm L}_i\widetilde{\bm x}_i - {\bm U}_i\widetilde{\bm x}_{i+1}\right]
    \label{eqn:xi}
\end{equation}
Following the logic of the cyclic reduction, Equation \ref{eqn:xi} can be used to eliminate ${\bm x}_{i-1}$ and ${\bm x}_i$ in Equation \ref{eqn:system1}, which forms the sub-system in Equation \ref{eqn:sub-system}.
\begin{equation}
    \widehat{\bm L}_i \widetilde{\bm x}_{i-1} + \widehat{\bm D}_i \widetilde{\bm x}_i + \widehat{\bm U}_i \widetilde{\bm x}_{i+1} = \widehat{\bm b}_i
    \label{eqn:sub-system}
\end{equation}
where
\begin{align}
    \widehat{\bm L}_i &= -\widetilde{\bm L}_i{\bm D}_{i-1}^{-1}{\bm L}_{i-1}
    \label{eqn:Li_hat_def}
    \\
    \widehat{\bm D}_i &= \widetilde{\bm D}_i - \widetilde{\bm L}_i{\bm D}_{i-1}^{-1}{\bm U}_{i-1} - \widetilde{\bm U}_i{\bm D}_i^{-1}{\bm L}_i
    \label{eqn:Di_hat_def}
    \\
    \widehat{\bm U}_i &= -\widetilde{\bm U}_i{\bm D}_i^{-1}{\bm U}_i
    \label{eqn:Ui_hat_def}
    \\
    \widehat{\bm b}_i &= \widetilde{\bm b}_i - \widetilde{\bm L}_i{\bm D}_{i-1}^{-1}{\bm b}_{i-1} - \widetilde{\bm U}_i{\bm D}_i^{-1}{\bm b}_i
    \label{eqn:bi_hat_def}
\end{align}
Equation \ref{eqn:sub-system} can be represented as $\widehat{\bm A}\widetilde{\bm x} = \widehat{\bm b}$, where $\widehat{\bm A}$ is a block tridiagonal system. If $\bm A$ is cyclic, then $\widehat{\bm A}$ is also cyclic. Considering the grid decomposition strategy, each block in $\widetilde{\bm x}_i$ or $\widehat{\bm b}_i$ is stored across the distributed memory, and each block can be solved efficiently with PCR. This data storage pattern is favorable for PCR, because the blocks can be easily located by the rank of the aligned grid decomposition to conduct the data transfer across the distributed memory. Once the sub-system is solved, all the $\widetilde{\bm x}_i$ are known, and the results can be propagated backward to solve ${\bm x}_i$ in parallel.

\begin{figure}[htbp]
    \centering
    \subfloat[]
    {\scalebox{0.3}{\tikzset{
    thick/.style={line width=2.5pt},
    thin/.style={line width=1.0pt},
    block d/.pic={
        \draw[fill=magenta!50!white, draw=black, thick] (0,0) rectangle (1,-1);
    },
    block l/.pic={
        \fill[magenta!50!white] (0,0) -- (1,-1) |- cycle;
        \draw[draw=black, thick] (-2,0) rectangle (1,-1);
    },
    block u/.pic={
        \fill[magenta!50!white] (0,0) -- (1,-1) -| cycle;
        \draw[draw=black, thick] (0,0) rectangle (3,-1);
    },
    block D/.pic={
        \fill[magenta!50!white] (0,0) -- (1,0) -- (3,-2) -- (3,-3) -- (2,-3) -- (0,-1) -- cycle;
        \draw[draw=black, thick] (0,0) rectangle (3,-3);
    },
    block L/.pic={
        \fill[magenta!50!white] (0,0) -- (1,-1) |- cycle;
        \draw[draw=black, thick] (0,0) rectangle (1,-3);
    },
    block U/.pic={
        \fill[magenta!50!white] (0,-2) -- (1,-3) -| cycle;
        \draw[draw=black, thick] (0,0) rectangle (1,-3);
    },
}

\begin{tikzpicture}
    \draw[draw=black, thick] (0,0) rectangle (16,-16);
    \foreach \i in {1,...,3} {
        \draw[draw=black, thin, dashed] (\i*4,0) -- (\i*4,-16);
        \draw[draw=black, thin, dashed] (0,-\i*4) -- (16,-\i*4);
    }
    \foreach \i in {0,...,3} {
        \pic at (\i*4, -\i*4) {block d};
        \pic at (\i*4+1, -\i*4) {block u};
        \pic at (\i*4+1, -\i*4-1) {block D};
        \pic at (\i*4, -\i*4-1) {block L};
        \ifnum \i>0
            \pic at (\i*4-1, -\i*4) {block l};
        \else
            \pic at (\i*4-1+16, -\i*4) {block l};
        \fi
        \ifnum \i<3
            \pic at (\i*4+4, -\i*4-1) {block U};
        \else
            \pic at (\i*4+4-16, -\i*4-1) {block U};
        \fi
    }
\end{tikzpicture}}\label{fig:BLU0}}
    \qquad
    \subfloat[]
    {\scalebox{0.3}{\tikzset{
    thick/.style={line width=2.5pt},
    thin/.style={line width=1.0pt},
    block d/.pic={
        \draw[fill=magenta!50!white, draw=black, thick] (0,0) rectangle (1,-1);
    },
    block l/.pic={
        \fill[magenta!50!white] (0,0) -- (1,-1) |- cycle;
        \draw[draw=black, thick] (-2,0) rectangle (1,-1);
    },
    block u/.pic={
        \fill[magenta!50!white] (0,0) -- (1,-1) -| cycle;
        \draw[draw=black, thick] (0,0) rectangle (3,-1);
    },
    block D/.pic={
        \fill[magenta!50!white] (0,0) -- (1,0) -- (3,-2) -- (3,-3) -- (2,-3) -- (0,-1) -- cycle;
        \draw[draw=black, thick] (0,0) rectangle (3,-3);
    },
    block L/.pic={
        \fill[magenta!50!white] (0,0) -- (1,-1) |- cycle;
        \draw[draw=black, thick] (0,0) rectangle (1,-3);
    },
    block U/.pic={
        \fill[magenta!50!white] (0,-2) -- (1,-3) -| cycle;
        \draw[draw=black, thick] (0,0) rectangle (1,-3);
    },
}

\begin{tikzpicture}
    \draw[draw=black, thick] (0,0) rectangle (16,-16);
    \foreach \i in {1,...,3} {
        \draw[draw=black, thin, dashed] (\i*3,0) -- (\i*3,-16);
        \draw[draw=black, thin, dashed] (0,-\i*3) -- (16,-\i*3);
        \draw[draw=black, thin, dashed] (\i+12,0) -- (\i+12,-16);
        \draw[draw=black, thin, dashed] (0,-\i-12) -- (16,-\i-12);
    }
    \draw[draw=black, thick] (12,0) -- (12,-16);
    \draw[draw=black, thick] (0,-12) -- (16,-12);
    \foreach \i in {0,...,3} {
        \pic at (\i+12, -\i-12) {block d};
        \pic at (\i*3, -\i*3) {block D};
        \pic at (\i*3, -\i-12) {block u};
        \pic at (\i+12, -\i*3) {block L};
        \ifnum \i>0
            \pic at (\i*3-1, -\i-12) {block l};
        \else
            \pic at (\i*3-1+12, -\i-12) {block l};
        \fi
        \ifnum \i<3
            \pic at (\i+12+1, -\i*3) {block U};
        \else
            \pic at (\i+12+1-4, -\i*3) {block U};
        \fi
    }
\end{tikzpicture}}\label{fig:BLU1}}
    \qquad
    \subfloat[]
    {\scalebox{0.3}{\tikzset{
    thick/.style={line width=2.5pt},
    thin/.style={line width=1.0pt},
    block d/.pic={
        \draw[fill=blue!70!white, draw=black, thick] (0,0) rectangle (1,-1);
    },
    block l/.pic={
        \draw[fill=green!30!white, draw=black, thick] (0,0) rectangle (1,-1);
    },
    block u/.pic={
        \draw[fill=yellow!80!white, draw=black, thick] (0,0) rectangle (1,-1);
    },
    block D/.pic={
        \fill[magenta!50!white] (0,0) -- (1,0) -- (3,-2) -- (3,-3) -- (2,-3) -- (0,-1) -- cycle;
        \draw[draw=black, thick] (0,0) rectangle (3,-3);
    },
    block L/.pic={
        \fill[magenta!50!white] (0,0) -- (1,-1) |- cycle;
        \draw[draw=black, thick] (0,0) rectangle (1,-3);
    },
    block U/.pic={
        \fill[magenta!50!white] (0,-2) -- (1,-3) -| cycle;
        \draw[draw=black, thick] (0,0) rectangle (1,-3);
    },
}

\begin{tikzpicture}
    \draw[draw=black, thick] (0,0) rectangle (16,-16);
    \foreach \i in {1,...,3} {
        \draw[draw=black, thin, dashed] (\i*3,0) -- (\i*3,-16);
        \draw[draw=black, thin, dashed] (0,-\i*3) -- (16,-\i*3);
        \draw[draw=black, thin, dashed] (\i+12,0) -- (\i+12,-16);
        \draw[draw=black, thin, dashed] (0,-\i-12) -- (16,-\i-12);
    }
    \draw[draw=black, thick] (12,0) -- (12,-16);
    \draw[draw=black, thick] (0,-12) -- (16,-12);
    \foreach \i in {0,...,3} {
        \pic at (\i+12, -\i-12) {block d};
        \pic at (\i*3, -\i*3) {block D};
        \pic at (\i+12, -\i*3) {block L};
        \ifnum \i>0
            \pic at (\i+12-1, -\i-12) {block l};
        \else
            \pic at (\i+12-1+4, -\i-12) {block l};
        \fi
        \ifnum \i<3
            \pic at (\i+12+1, -\i*3) {block U};
            \pic at (\i+12+1, -\i-12) {block u};
        \else
            \pic at (\i+12+1-4, -\i*3) {block U};
            \pic at (\i+12+1-4, -\i-12) {block u};
        \fi
    }
\end{tikzpicture}}\label{fig:BLU2}}
    \caption{Sparsity patterns of the system during permutation and block LU-factorization. \protect\subref{fig:BLU0} is the original matrix ${\bm A}$; \protect\subref{fig:BLU1} is the permuted matrix ${\bm P}{\bm A}{\bm P}^T$; and \protect\subref{fig:BLU2} is the block upper triangular matrix obtained via the block LU-factorization from ${\bm P}{\bm A}{\bm P}^T$.}
    \label{fig:BLU}
\end{figure}
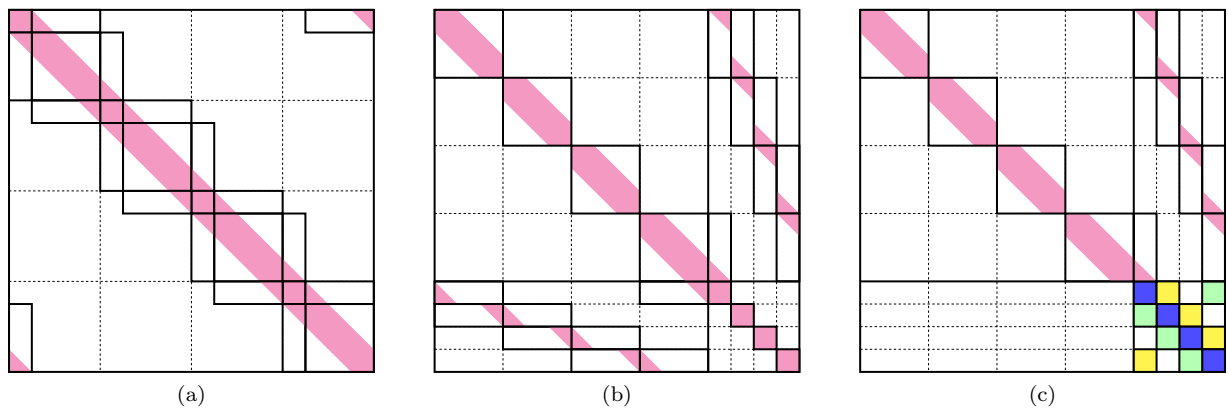

The method can be also interpreted as a block LU-factorization, analogous to the illustration in Gander and Golub \cite{gander1997cyclic}. Introducing a permutation matrix ${\bm P}$, the linear system, ${\bm A}{\bm x} = {\bm b}$, can be modified to $({\bm P}{\bm A}{\bm P}^T)({\bm P}{\bm x}) = {\bm P}{\bm b}$, where the row and column permutations, ${\bm P}{\bm A}{\bm P}^T$, regroup ${\bm D}_i$ and $\widetilde{\bm D}_i$ respectively. The resulting pattern is shown in Figure \ref{fig:BLU1}. The ${\bm D}_i$ blocks remain in the top left region on the diagonal, and the $\widetilde{\bm D}_i$ blocks are moved to the bottom right region also on the diagonal. Correspondingly, the $\widetilde{\bm L}_i$ and $\widetilde{\bm U}_i$ blocks show up in the bottom left region, and ${\bm L}_i$ and ${\bm U}_i$ blocks are placed in the top right region. The process to obtain Equation \ref{eqn:sub-system} is block Gaussian elimination. As a result, the permuted system becomes a block upper triangular system as shown in Figure \ref{fig:BLU2}, and the sub-system $\widehat{\bm A}$ is formed as the last diagonal block. Additionally, it is clearly shown in Figure \ref{fig:BLU2} that the top left region only contains the diagonal located blocks, ${\bm D}_i$. All the non-diagonal blocks are coupled ${\bm D}_i$ with $\widehat{\bm A}$ only, and no coupling is created among different ${\bm D}_i$ blocks. This reaffirms that once the sub-system, $\widehat{\bm A}\widetilde{x} = \widehat{\bm b}$, is solved, then the remaining sub-system, formed by Equation \ref{eqn:system2}, can be solved in parallel on each data partition.

The following section discusses the solution method of the sub-system, $\widehat{\bm A}\widetilde{\bm x} = \widehat{\bm b}$. As aforementioned, $\widehat{\bm A}$ is a block tridiagonal system, which may be cyclic depending on the original banded system, $\bm A$. The block size depends on the half band width of ${\bm A}$, and the dimension of ${\widehat{\bm A}}$ equals the number of the aligned grid partitions. The ``dimension'' of $\widehat{\bm A}$ refers to the number of blocks in each row and column in $\widehat{\bm A}$. Each block in $\widetilde{\bm x}$ and $\widehat{\bm b}$ are stored in a unique partition. 
With non-periodic boundaries, $\widehat{\bm A}$ is acyclic, and the solution method will follow the block PCR in a fairly straightforward way. With periodic boundaries, $\widehat{\bm A}$ is cyclic, so a non-zero block will show up in the top right and bottom left corners. In this case, if the dimension of $\widehat{\bm A}$ is a power of two, PCR can be directly applied. PCR can still be applied for cyclic $\widehat{\bm A}$ of arbitrary dimension using special treatment. Sweet, in his work \cite{sweet1977cyclic}, suggests such a treatment for cyclic block tridiagonal systems. However, considering the complexity of data storage and data migration, a different treatment is proposed in this paper which requires the dimension of a sub-system of $\widehat{\bm A}$ undergoing a PCR step to be even. If the dimension is odd, a detaching step is needed before the PCR step. During the detaching step, the last row of each sub-system will be used to eliminate the upper and lower off-diagonal blocks of the previous row and the first row of the same sub-system respectively, and then detached from the sub-system. For periodicity, the lower diagonal block in the first row is placed in the last column. After this step, the dimension of each sub-system is a even number, which is ready for the next PCR step. The detached rows will then be addressed and reattached to the sub-system through a backward substitution phase after the rows are solved.

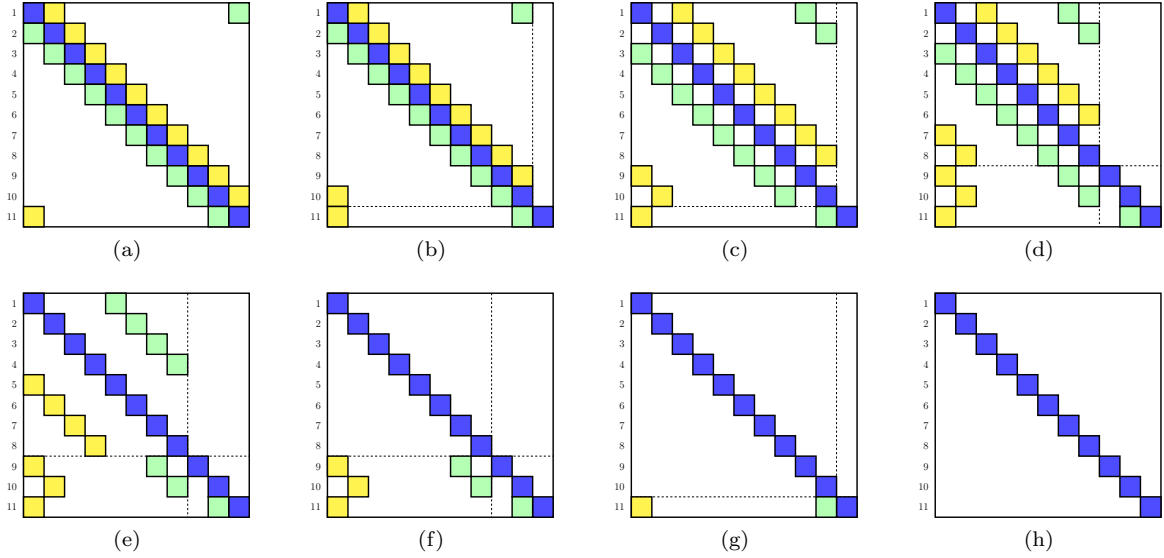
\begin{figure}[htbp]
    \centering
    \subfloat[]{\scalebox{0.27}{\tikzset{
    thick/.style={line width=2.0pt},
    thin/.style={line width=1.0pt},
    diagonal block/.pic={
        \draw[fill=blue!70!white, draw=black, thick] (0,0) rectangle (1,-1);
    },
    upper block/.pic={
        \draw[fill=yellow!80!white, draw=black, thick] (0,0) rectangle (1,-1);
    },
    lower block/.pic={
        \draw[fill=green!30!white, draw=black, thick] (0,0) rectangle (1,-1);
    },
}

\begin{tikzpicture}
    \draw[draw=black, thick] (0,0) rectangle (11,-11);
    \foreach \i in {0,...,10} {
        \pic at (\i,-\i) {diagonal block};
    }
    \foreach \i in {0,...,10} {
        \pgfmathparse{\i<10 ? 1 : 0}
            \ifnum \pgfmathresult>0
                \pic at (\i+1,-\i) {upper block};
            \else
                \pic at (\i+1-11,-\i) {upper block};
            \fi
        \pgfmathparse{\i>0 ? 1 : 0}
            \ifnum \pgfmathresult>0
                \pic at (\i-1,-\i) {lower block};
            \else
                \pic at (\i-1+11,-\i) {lower block};
            \fi
    }
    \foreach \i in {1,...,11} {
        \node [anchor=east] at (-0.2, -\i+0.5) {\Large\i};
    }
\end{tikzpicture}}\label{fig:period_pcr_0}}
    \qquad
    \subfloat[]{\scalebox{0.27}{\tikzset{
    thick/.style={line width=2.0pt},
    thin/.style={line width=1.0pt},
    diagonal block/.pic={
        \draw[fill=blue!70!white, draw=black, thick] (0,0) rectangle (1,-1);
    },
    upper block/.pic={
        \draw[fill=yellow!80!white, draw=black, thick] (0,0) rectangle (1,-1);
    },
    lower block/.pic={
        \draw[fill=green!30!white, draw=black, thick] (0,0) rectangle (1,-1);
    },
}

\begin{tikzpicture}
    \draw[draw=black, thick] (0,0) rectangle (11,-11);
    \draw[draw=black, dashed, thin] ( 0,-10) -- (11,-10);
    \draw[draw=black, dashed, thin] (10,  0) -- (10,-11);
    \foreach \i in {0,...,10} {
        \pic at (\i,-\i) {diagonal block};
    }
    \foreach \i in {0,...,9} {
        \pgfmathparse{\i+1<10 ? 1 : 0}
            \ifnum \pgfmathresult>0
                \pic at (\i+1,-\i) {upper block};
            \else
                \pic at (\i+1-10,-\i) {upper block};
            \fi
        \pgfmathparse{\i-1>=0 ? 1 : 0}
            \ifnum \pgfmathresult>0
                \pic at (\i-1,-\i) {lower block};
            \else
                \pic at (\i-1+10,-\i) {lower block};
            \fi
    }
    \pic at (0,-10) {upper block};
    \pic at (9,-10) {lower block};
    
    \foreach \i in {1,...,11} {
        \node [anchor=east] at (-0.2, -\i+0.5) {\Large\i};
    }
\end{tikzpicture}}\label{fig:period_pcr_1}}
    \qquad
    \subfloat[]{\scalebox{0.27}{\tikzset{
    thick/.style={line width=2.0pt},
    thin/.style={line width=1.0pt},
    diagonal block/.pic={
        \draw[fill=blue!70!white, draw=black, thick] (0,0) rectangle (1,-1);
    },
    upper block/.pic={
        \draw[fill=yellow!80!white, draw=black, thick] (0,0) rectangle (1,-1);
    },
    lower block/.pic={
        \draw[fill=green!30!white, draw=black, thick] (0,0) rectangle (1,-1);
    },
}

\begin{tikzpicture}
    \draw[draw=black, thick] (0,0) rectangle (11,-11);
    \draw[draw=black, dashed, thin] ( 0,-10) -- (11,-10);
    \draw[draw=black, dashed, thin] (10,  0) -- (10,-11);
    \foreach \i in {0,...,10} {
        \pic at (\i,-\i) {diagonal block};
    }
    \foreach \i in {0,...,9} {
        \pgfmathparse{\i+2<10 ? 1 : 0}
            \ifnum \pgfmathresult>0
                \pic at (\i+2,-\i) {upper block};
            \else
                \pic at (\i+2-10,-\i) {upper block};
            \fi
        \pgfmathparse{\i-2>=0 ? 1 : 0}
            \ifnum \pgfmathresult>0
                \pic at (\i-2,-\i) {lower block};
            \else
                \pic at (\i-2+10,-\i) {lower block};
            \fi
    }
    \pic at (0,-10) {upper block};
    \pic at (9,-10) {lower block};
    
    \foreach \i in {1,...,11} {
        \node [anchor=east] at (-0.2, -\i+0.5) {\Large\i};
    }
\end{tikzpicture}}\label{fig:period_pcr_2}}
    \qquad
    \subfloat[]{\scalebox{0.27}{\tikzset{
    thick/.style={line width=2.0pt},
    thin/.style={line width=1.0pt},
    diagonal block/.pic={
        \draw[fill=blue!70!white, draw=black, thick] (0,0) rectangle (1,-1);
    },
    upper block/.pic={
        \draw[fill=yellow!80!white, draw=black, thick] (0,0) rectangle (1,-1);
    },
    lower block/.pic={
        \draw[fill=green!30!white, draw=black, thick] (0,0) rectangle (1,-1);
    },
}

\begin{tikzpicture}
    \draw[draw=black, thick] (0,0) rectangle (11,-11);
    \draw[draw=black, dashed, thin] (0,-8) -- (11,-8);
    \draw[draw=black, dashed, thin] (8, 0) -- (8,-11);
    \foreach \i in {0,...,10} {
        \pic at (\i,-\i) {diagonal block};
    }
    \foreach \i in {0,...,7} {
        \pgfmathparse{\i+2<8 ? 1 : 0}
            \ifnum \pgfmathresult>0
                \pic at (\i+2,-\i) {upper block};
            \else
                \pic at (\i+2-8,-\i) {upper block};
            \fi
        \pgfmathparse{\i-2>=0 ? 1 : 0}
            \ifnum \pgfmathresult>0
                \pic at (\i-2,-\i) {lower block};
            \else
                \pic at (\i-2+8,-\i) {lower block};
            \fi
    }
    \pic at (0,-10) {upper block};
    \pic at (9,-10) {lower block};
    \pic at (0, -8) {upper block};
    \pic at (1, -9) {upper block};
    \pic at (6, -8) {lower block};
    \pic at (7, -9) {lower block};
    
    \foreach \i in {1,...,11} {
        \node [anchor=east] at (-0.2, -\i+0.5) {\Large\i};
    }
\end{tikzpicture}}\label{fig:period_pcr_3}}
    \\
    \subfloat[]{\scalebox{0.27}{\tikzset{
    thick/.style={line width=2.0pt},
    thin/.style={line width=1.0pt},
    diagonal block/.pic={
        \draw[fill=blue!70!white, draw=black, thick] (0,0) rectangle (1,-1);
    },
    upper block/.pic={
        \draw[fill=yellow!80!white, draw=black, thick] (0,0) rectangle (1,-1);
    },
    lower block/.pic={
        \draw[fill=green!30!white, draw=black, thick] (0,0) rectangle (1,-1);
    },
    overlapped block/.pic={
        \draw[fill=lime!80!white, draw=black, thick] (0,0) rectangle (1,-1);
    },
}

\begin{tikzpicture}
    \draw[draw=black, thick] (0,0) rectangle (11,-11);
    \draw[draw=black, dashed, thin] (0,-8) -- (11,-8);
    \draw[draw=black, dashed, thin] (8, 0) -- (8,-11);
    \foreach \i in {0,...,10} {
        \pic at (\i,-\i) {diagonal block};
    }
    \foreach \i in {0,...,7} {
        \pgfmathparse{\i+4<8 ? 1 : 0}
        \ifnum \pgfmathresult>0
            \pic at (\i+4,-\i) {lower block};
        \else
            \pic at (\i+4-8,-\i) {upper block};
        \fi
    }
    \pic at (0,-10) {upper block};
    \pic at (9,-10) {lower block};
    \pic at (0, -8) {upper block};
    \pic at (1, -9) {upper block};
    \pic at (6, -8) {lower block};
    \pic at (7, -9) {lower block};
    
    \foreach \i in {1,...,11} {
        \node [anchor=east] at (-0.2, -\i+0.5) {\Large\i};
    }
\end{tikzpicture}}\label{fig:period_pcr_4}}
    \qquad
    \subfloat[]{\scalebox{0.27}{\tikzset{
    thick/.style={line width=2.0pt},
    thin/.style={line width=1.0pt},
    diagonal block/.pic={
        \draw[fill=blue!70!white, draw=black, thick] (0,0) rectangle (1,-1);
    },
    upper block/.pic={
        \draw[fill=yellow!80!white, draw=black, thick] (0,0) rectangle (1,-1);
    },
    lower block/.pic={
        \draw[fill=green!30!white, draw=black, thick] (0,0) rectangle (1,-1);
    },
    overlapped block/.pic={
        \draw[fill=lime!80!white, draw=black, thick] (0,0) rectangle (1,-1);
    },
}

\begin{tikzpicture}
    \draw[draw=black, thick] (0,0) rectangle (11,-11);
    \draw[draw=black, dashed, thin] (0,-8) -- (11,-8);
    \draw[draw=black, dashed, thin] (8, 0) -- (8,-11);
    \foreach \i in {0,...,10} {
        \pic at (\i,-\i) {diagonal block};
    }
    \pic at (0,-10) {upper block};
    \pic at (9,-10) {lower block};
    \pic at (0, -8) {upper block};
    \pic at (1, -9) {upper block};
    \pic at (6, -8) {lower block};
    \pic at (7, -9) {lower block};
    \foreach \i in {1,...,11} {
        \node [anchor=east] at (-0.2, -\i+0.5) {\Large\i};
    }
\end{tikzpicture}}\label{fig:period_pcr_5}}
    \qquad
    \subfloat[]{\scalebox{0.27}{\tikzset{
    thick/.style={line width=2.0pt},
    thin/.style={line width=1.0pt},
    diagonal block/.pic={
        \draw[fill=blue!70!white, draw=black, thick] (0,0) rectangle (1,-1);
    },
    upper block/.pic={
        \draw[fill=yellow!80!white, draw=black, thick] (0,0) rectangle (1,-1);
    },
    lower block/.pic={
        \draw[fill=green!30!white, draw=black, thick] (0,0) rectangle (1,-1);
    },
    overlapped block/.pic={
        \draw[fill=lime!80!white, draw=black, thick] (0,0) rectangle (1,-1);
    },
}

\begin{tikzpicture}
    \draw[draw=black, thick] (0,0) rectangle (11,-11);
    \draw[draw=black, dashed, thin] ( 0,-10) -- (11,-10);
    \draw[draw=black, dashed, thin] (10,  0) -- (10,-11);
    \foreach \i in {0,...,10} {
        \pic at (\i,-\i) {diagonal block};
    }
    \pic at (0,-10) {upper block};
    \pic at (9,-10) {lower block};
    \foreach \i in {1,...,11} {
        \node [anchor=east] at (-0.2, -\i+0.5) {\Large\i};
    }
\end{tikzpicture}}\label{fig:period_pcr_6}}
    \qquad
    \subfloat[]{\scalebox{0.27}{\tikzset{
    thick/.style={line width=2.0pt},
    thin/.style={line width=1.0pt},
    diagonal block/.pic={
        \draw[fill=blue!70!white, draw=black, thick] (0,0) rectangle (1,-1);
    },
    upper block/.pic={
        \draw[fill=yellow!80!white, draw=black, thick] (0,0) rectangle (1,-1);
    },
    lower block/.pic={
        \draw[fill=green!30!white, draw=black, thick] (0,0) rectangle (1,-1);
    },
    overlapped block/.pic={
        \draw[fill=lime!80!white, draw=black, thick] (0,0) rectangle (1,-1);
    },
}

\begin{tikzpicture}
    \draw[draw=black, thick] (0,0) rectangle (11,-11);
    \foreach \i in {0,...,10} {
        \pic at (\i,-\i) {diagonal block};
    }
    \foreach \i in {1,...,11} {
        \node [anchor=east] at (-0.2, -\i+0.5) {\Large\i};
    }
\end{tikzpicture}}\label{fig:period_pcr_7}}
    \caption{Reduction procedure of an $11\times11$ $\widehat{\bm A}$. From \protect\subref{fig:period_pcr_0} to \protect\subref{fig:period_pcr_1}, the row 11 is detached from the sub-system; from \protect\subref{fig:period_pcr_1} to \protect\subref{fig:period_pcr_2}, two sub-systems are formed by a PCR step; from \protect\subref{fig:period_pcr_2} to \protect\subref{fig:period_pcr_3}, row 9 and row 10 are detached from the sub-systems; from \protect\subref{fig:period_pcr_3} to \protect\subref{fig:period_pcr_5}, all the eight unknowns in the sub-systems are solved; from \protect\subref{fig:period_pcr_5} to \protect\subref{fig:period_pcr_6}, solutions backward propagate to the first level to solve row 9 and row 10; from \protect\subref{fig:period_pcr_6} to \protect\subref{fig:period_pcr_7}, solutions backwards propagate to the root level to solve row 11.}
    \label{fig:subsystem_sparsity}
\end{figure}

An example is provided by setting $\widehat{\bm A}$ to be a $11\times11$ cyclic tridiagonal matrix. The sparsity pattern in each step is visualized in Figure \ref{fig:subsystem_sparsity}, and the communication pattern is shown in Figure \ref{fig:modified_pcr_tree}. On the root level, the number of sub-systems is $1$, and the dimension is $11$. Since the dimension of this subsystem is odd, the last row needs to detach from the sub-system before conducting PCR. Use the last row to eliminate the upper off-diagonal element of the tenth row and the lower off-diagonal element of the first row, so that a $10\times10$ sub-system is created and the last row is detached, as shown in Figure \ref{fig:period_pcr_1}. After a PCR step, the $10\times10$ sub-system is split into two $5\times5$ sub-system on the first level, as shown in Figure \ref{fig:period_pcr_2}. Before conducting PCR on the first level, the last row of each of the two sub-systems (row 9 and row 10) needs to be detached. Row 9 is used to eliminate the upper off-diagonal element of row 7 (the second to last row of its sub-system on this level) and the lower off-diagonal element of row 1. Row 10 is used to eliminate the upper off-diagonal element of row 8 and the lower diagonal element of row 2 (the first row of its sub-system on this level), so two sub-systems are reduced to $4\times4$ as shown in Figure \ref{fig:period_pcr_3}. Starting from this level, the number of rows involved in the remaining PCR steps is eight, which is a power of two. At this point, no further detachment is needed, and all the eight unknowns can be solved by two steps of PCR. Then, the eight solutions are backwards substituted into the two $5\times5$ sub-systems on the first level to solve the row 9 and row 10. In the final step, the ten solutions propagate backwards to the root level, and are substituted into the $11\times11$ system to solve row 11, so that all the unknowns are solved.

\begin{figure}[htbp]
    \centering
    \scalebox{0.4}{\tikzset{
    thick/.style={line width=2.5pt},
    thin/.style={line width=1.5pt},
}
\begin{tikzpicture}
    \foreach \i in {1,...,11} {
        \draw[thick, fill=green!20!white] (\i*2, 0) circle (0.7);
        \node [anchor=center] at (\i*2, 0) {\LARGE\i};
    }
    \draw[->, >=stealth, thin] (11*2, -1) to[out=-20, in=160] ++ (-10*2,-2);
    \draw[->, >=stealth, thin] (11*2, -1) --++ ( -1*2,-2) ;
    \foreach \i in {1,...,10} {
        \draw[->, >=stealth, thin] (\i*2, -1) --++ (0,-2);
        \draw[thick, fill=green!20!white] (\i*2,-4) circle (0.7);
        \node [anchor=center] at (\i*2,-4) {\LARGE\i};
    }
    
    \foreach \i in {1,...,5} {
        \draw[thick, fill=yellow!50!white] (\i*4-2,-8) circle (0.7);
        \draw[thick, fill=  blue!40!white] (\i*4,  -8) circle (0.7);
        \foreach \j in {0,...,1} {
            \pgfmathparse{\i*2-\j>1 ? 1 : 0}
            \ifnum \pgfmathresult>0
                \draw[->, >=stealth, thin] (\i*4-\j*2, -5) --++ (-2,-2);
            \fi
            \pgfmathparse{\i*2-\j<10 ? 1 : 0}
            \ifnum \pgfmathresult>0
                \draw[->, >=stealth, thin] (\i*4-\j*2, -5) --++ ( 2,-2);
            \fi
        }
    }
    \draw[->, >=stealth, thin] (10*2, -5) to[out=-20, in=160] ++ (-9*2,-2);
    \draw[->, >=stealth, thin] (   2, -5) to[out=-160,in= 20] ++ ( 9*2,-2);
    \foreach \i in {1,...,10} {
        \draw[->, >=stealth, thin] (\i*2, -5) --++ (0,-2);
        \node [anchor=center] at (\i*2,-8) {\LARGE\i};
    }
    
    \foreach \i in {1,...,4} {
        \draw[thick, fill=yellow!50!white] (\i*4-2,-12) circle (0.7);
        \draw[thick, fill=  blue!40!white] (\i*4,  -12) circle (0.7);
    }
    \foreach \i in {1,...,8} {
        \draw[->, >=stealth, thin] (\i*2, -9) --++ (0,-2);
        \node [anchor=center] at (\i*2,-12) {\LARGE\i};
    }
    \draw[->, >=stealth, thin] ( 9*2, -9) --++ (-4,-2);
    \draw[->, >=stealth, thin] (10*2, -9) --++ (-4,-2);
    \draw[->, >=stealth, thin] (10*2, -9) to[out=-20, in=160] ++ (-8*2,-2);
    \draw[->, >=stealth, thin] ( 9*2, -9) to[out=-20, in=160] ++ (-8*2,-2);
    
    \foreach \i in {1,...,2} {
        \draw[thick, fill=purple!50!white] (\i*8-6,-16) circle (0.7);
        \draw[thick, fill=  cyan!10!white] (\i*8-4,-16) circle (0.7);
        \draw[thick, fill=  lime!60!white] (\i*8-2,-16) circle (0.7);
        \draw[thick, fill= brown!30!white] (\i*8,  -16) circle (0.7);
        \foreach \j in {0,...,3} {
            \pgfmathparse{\i*4-\j>2 ? 1 : 0}
            \ifnum \pgfmathresult>0
                \draw[->, >=stealth, thin] (\i*8-\j*2, -13) --++ (-4,-2);
            \fi
            \pgfmathparse{\i*4-\j<7 ? 1 : 0}
            \ifnum \pgfmathresult>0
                \draw[->, >=stealth, thin] (\i*8-\j*2, -13) --++ ( 4,-2);
            \fi
        }
    }
    \draw[->, >=stealth, thin] (8*2,-13) to[out=-20, in=160] ++ (-6*2,-2);
    \draw[->, >=stealth, thin] (7*2,-13) to[out=-20, in=160] ++ (-6*2,-2);
    \draw[->, >=stealth, thin] (1*2,-13) to[out=-160, in=20] ++ ( 6*2,-2);
    \draw[->, >=stealth, thin] (2*2,-13) to[out=-160, in=20] ++ ( 6*2,-2);
    \foreach \i in {1,...,8} {
        \draw[->, >=stealth, thin] (\i*2, -13) --++ (0,-2);
        \node [anchor=center] at (\i*2,-16) {\LARGE\i};
    }
    
    \foreach \i in {1,...,2} {
        \draw[thick, fill=purple!50!white] (\i*8-6,-16) circle (0.7);
        \draw[thick, fill=  cyan!10!white] (\i*8-4,-16) circle (0.7);
        \draw[thick, fill=  lime!60!white] (\i*8-2,-16) circle (0.7);
        \draw[thick, fill= brown!30!white] (\i*8,  -16) circle (0.7);
        \foreach \j in {0,...,3} {
            \pgfmathparse{\i*4-\j>2 ? 1 : 0}
            \ifnum \pgfmathresult>0
                \draw[->, >=stealth, thin] (\i*8-\j*2, -13) --++ (-4,-2);
            \fi
            \pgfmathparse{\i*4-\j<7 ? 1 : 0}
            \ifnum \pgfmathresult>0
                \draw[->, >=stealth, thin] (\i*8-\j*2, -13) --++ ( 4,-2);
            \fi
        }
    }
    \draw[->, >=stealth, thin] (8*2,-13) to[out=-20, in=160] ++ (-6*2,-2);
    \draw[->, >=stealth, thin] (7*2,-13) to[out=-20, in=160] ++ (-6*2,-2);
    \draw[->, >=stealth, thin] (1*2,-13) to[out=-160, in=20] ++ ( 6*2,-2);
    \draw[->, >=stealth, thin] (2*2,-13) to[out=-160, in=20] ++ ( 6*2,-2);
    \foreach \i in {1,...,8} {
        \draw[->, >=stealth, thin] (\i*2, -13) --++ (0,-2);
        \node [anchor=center] at (\i*2,-16) {\LARGE\i};
    }
    
    \draw[thick, fill=  teal!50!white] ( 2,-20) circle (0.7);
    \draw[thick, fill= olive!50!white] ( 4,-20) circle (0.7);
    \draw[thick, fill=orange!70!white] ( 6,-20) circle (0.7);
    \draw[thick, fill=yellow!90!white] ( 8,-20) circle (0.7);
    \draw[thick, fill=purple!20!white] (10,-20) circle (0.7);
    \draw[thick, fill=  cyan!30!white] (10,-20) circle (0.7);
    \draw[thick, fill=violet!40!white] (12,-20) circle (0.7);
    \draw[thick, fill= green!50!white] (14,-20) circle (0.7);
    \draw[thick, fill=  blue!20!white] (16,-20) circle (0.7);
    \foreach \j in {0,...,7} {
        \pgfmathparse{8-\j>4 ? 1 : 0}
        \ifnum \pgfmathresult>0
            \draw[->, >=stealth, thin] (16-\j*2, -17) --++ (-8,-2);
        \fi
        \pgfmathparse{8-\j<5 ? 1 : 0}
        \ifnum \pgfmathresult>0
            \draw[->, >=stealth, thin] (16-\j*2, -17) --++ ( 8,-2);
        \fi
    }
    \foreach \i in {1,...,8} {
        \draw[->, >=stealth, thin] (\i*2, -17) --++ (0,-2);
        \node [anchor=center] at (\i*2,-20) {\LARGE\i};
    }
    
    \foreach \i in {1,...,5} {
        \draw[thick, fill=yellow!50!white] (\i*4-2,-24) circle (0.7);
        \draw[thick, fill=  blue!40!white] (\i*4,  -24) circle (0.7);
    }
    \draw[->, >=stealth, thin] (8*2, -21) --++ (4,-2);
    \draw[->, >=stealth, thin] (7*2, -21) --++ (4,-2);
    \draw[->, >=stealth, thin] (1*2,-21) to[out=-160, in=20] ++ (8*2,-2);
    \draw[->, >=stealth, thin] (2*2,-21) to[out=-160, in=20] ++ (8*2,-2);
    \draw[->, >=stealth, thin] ( 9*2, -9) --++ (0,-14);
    \draw[->, >=stealth, thin] (10*2, -9) --++ (0,-14);
    \foreach \i in {1,...,10} {
        \ifnum \i<9
            \draw[->, >=stealth, thin] (\i*2, -21) --++ (0,-2);
        \fi
        \node [anchor=center] at (\i*2,-24) {\LARGE\i};
    }
    
    \foreach \i in {1,...,11} {
        \ifnum \i<11
            \draw[->, >=stealth, thin] (\i*2, -25) --++ (0,-2);
        \fi
        \draw[thick, fill=green!20!white] (\i*2,-28) circle (0.7);
        \node [anchor=center] at (\i*2,-28) {\LARGE\i};
    }
    \draw[->, >=stealth, thin] (10*2, -25) --++ (2,-2);
    \draw[->, >=stealth, thin] (1*2,-25) to[out=-160, in=20] ++ (10*2,-2);
    \draw[->, >=stealth, thin] (11*2, -1) --++ (0,-26);
\end{tikzpicture}}
    \caption{Communication pattern of PCR for an $11\times11$ cyclic tridiagonal system. The sub-systems in each step are grouped by the same colors.  } 
    \label{fig:modified_pcr_tree}
\end{figure}

\section{Implementation details}
\label{sec:ImplementationDetails}
The terms ${\bm D}_i^{-1}{\bm L}_i$, ${\bm D}_i^{-1}{\bm U}_i$, and ${\bm D}_i^{-1}{\bm b}_i$, in Equations (\ref{eqn:Li_hat_def} -- \ref{eqn:bi_hat_def}), are computed by solving the following linear systems.
\begin{align}
    {\bm D}_i {\bm S}_i &= {\bm L}_i \label{eqn:Si}
    \\
    {\bm D}_i {\bm R}_i &= {\bm U}_i \label{eqn:Ri}
    \\
    {\bm D}_i {\bm y}_i &= {\bm b}_i \label{eqn:yi}
\end{align}
for ${\bm S}_i$, ${\bm R}_i$, and ${\bm y}_i$, respectively.
Based on the proposed approach, ${\bm D}_i$ is an acyclic, compact banded matrix, and all the data on the right-hand-side and the unknowns to be solved are stored in the same partition. Therefore, generalized PCR can be used to further parallelize these solves. Using generalized PCR to solve ${\bm S}_i$, ${\bm R}_i$, and ${\bm y}_i$, the number of the parallel reduction steps for each system is $\lceil{\log_2 N_i}\rceil$, where $N_i$ is the dimension of ${\bm D}_i$. All the operations at this stage are conducted on the shared memory simultaneously on each partition. Substituting ${\bm S}_i$, ${\bm R}_i$, and ${\bm y}_i$ into Equations (\ref{eqn:Li_hat_def} -- \ref{eqn:bi_hat_def}), the reduced system -- Equation  (\ref{eqn:sub-system}) -- can be practically constructed according to the following equations.
\begin{align}
    \widehat{\bm L}_i &= -\widetilde{\bm L}_i{\bm S}_{i-1}
    \label{eqn:Li_hat}
    \\
    \widehat{\bm D}_i &= \widetilde{\bm D}_i - \widetilde{\bm L}_i{\bm R}_{i-1} - \widetilde{\bm U}_i{\bm S}_i
    \label{eqn:Di_hat}
    \\
    \widehat{\bm U}_i &= -\widetilde{\bm U}_i{\bm R}_i
    \label{eqn:Ui_hat}
    \\
    \widehat{\bm b}_i &= \widetilde{\bm b}_i - \widetilde{\bm L}_i{\bm y}_{i-1} - \widetilde{\bm U}_i{\bm y}_i
    \label{eqn:bi_hat}
\end{align}
Following the proposed approach to solve for $\widetilde{\bm x}_i$ and substituting into Equation (\ref{eqn:xi}), ${\bm x}_i$ can be obtained by the following operation.
\begin{equation}
    {\bm x}_i = {\bm y}_i - {\bm S}_i\widetilde{\bm x}_i - {\bm R}_i\widetilde{\bm x}_{i+1}
    \label{eqn:xi_app}
\end{equation}
A sample implementation is shown in Algorithm \ref{alg:main} where the detaching step, block PCR step, and Reattaching step is shown in Algorithm \ref{alg:detach}, \ref{alg:blockPCR}, and \ref{alg:reattach}, respectively. The sample code is given in the MPI (message passing interface) style where the rank of partition starts from zero.

\begin{algorithm}[htbp]
    \SetKwData{sdata}{send\_buffer}
    \SetKwData{drank}{dest\_rank}
    \SetKwData{srank}{src\_rank}
    \SetKwData{tag}{tag}
    \SetKwData{tagA}{tag\_a}
    \SetKwData{tagB}{tag\_b}
    \SetKwFunction{genPCR}{generalizedPCR}
    \SetKwFunction{getData}{getFromPartition}
    \SetKwFunction{sendData}{sendToPartition}
    \SetKwFunction{emptyStack}{initEmptyStack}
    \SetKwFunction{push}{stackPush}
    \SetKwFunction{pop}{stackPop}
    \SetKwFunction{isNotEmpty}{isNotEmpty}
    \SetKwInOut{Input}{in}
    \SetKwInOut{Output}{in/out}
    
    \Input{
        ${\bm D}_i$,
        $\widetilde{\bm L}_i$,
        $\widetilde{\bm U}_i$,
        $i$,
        $p$
    }
    \Output{
        ${\bm W}_i\leftarrow\widetilde{\bm D}_i$,
        ${\bm Y}_i\leftarrow{\bm L}_i$,
        ${\bm Z}_i\leftarrow{\bm U}_i$,
        ${\bm x}_i\leftarrow{\bm b}_i$,
        $\widetilde{\bm x}_i\leftarrow\widetilde{\bm b}_i$
    }
    \BlankLine
    \tcc{Factorization}
    ${\bm Y}_i$ $\leftarrow$ \genPCR{${\bm D}_i$, ${\bm Y}_i$}\;
    ${\bm Z}_i$ $\leftarrow$ \genPCR{${\bm D}_i$, ${\bm Z}_i$}\;
    \tag $\leftarrow$ \sendData{\sdata$=\{{\bm Y}_i, {\bm Z}_i\}$, \drank$=(i+1)\mod p$}\;
    $\widehat{\bm U}_i$ $\leftarrow$ $-\widetilde{\bm U}_i{\bm Z}_i$\;
    $\{{\bm Y}_i, {\bm Z}_i\}$ $\leftarrow$ \getData{\tag, \srank$=(p+i-1)\mod p$}\;
    $\widehat{\bm L}_i$ $\leftarrow$ $-\widetilde{\bm L}_i{\bm Y}_{i-1}$\;
    ${\bm W}_i$ $\leftarrow$ ${\bm W}_i - \widetilde{\bm L}_i{\bm Z}_{i-1} - \widetilde{\bm U}_i{\bm Y}_i$\;
    
    \BlankLine
    \tcc{Solve reduced system}
    ${\bm x}_i$ $\leftarrow$ \genPCR{${\bm D}_i$, ${\bm x}_i$}\;
    \tag $\leftarrow$ \sendData{$\sdata={\bm x}_i$, $\drank = (i+1)\mod p$}\;
    ${\bm x}_{i-1}$ $\leftarrow$ \getData{\tag, $\srank = (p+i-1)\mod p$}\;
    $\widetilde{\bm x}_i$ $\leftarrow$ $\widetilde{\bm x}_i - \widetilde{\bm L}_i{\bm x}_{i-1} - \widetilde{\bm U}_i{\bm x}_i$\;
    \BlankLine
    $s$ $\leftarrow1$; \tcp{stride as well as the number of sub-systems}
    $n_0$  $\leftarrow$ $p$; \tcp{size of each sub-system in the current PCR step}
    $n_a$ $\leftarrow$ $p$; \tcp{number of attached rows in the PCR step $n_a\equiv s\times n_0$}
    $\mathcal{S}$ $\leftarrow$ \emptyStack{}; \tcp{a stack of boolean}
    \While{$n_0>1$} {
        $\mathcal{S}$ $\leftarrow$ \push{$n_0\mod 2 > 0$}\;
        \If{$n_0\mod 2 > 0$}{
            $n_0$ $\leftarrow$ $n_0 - 1$\;
            $n_a$ $\leftarrow$ $n_a - s$\;
            \emph{Detach the last row of each sub-system (See Algorithm \ref{alg:detach})}\;
        }
        \emph{Block PCR step (See Algorithm \ref{alg:blockPCR})}\;

        $s$ $\leftarrow$ $s\times 2$\;
        $n_0$ $\leftarrow$ $n_0 / 2$\;
    } 
    \If{$i < n_a$}{
        $\widetilde{\bm x}_i$ $\leftarrow$ ${\bm W}_i^{-1}\widetilde{\bm x}_i$\;
    }    
    \While{\isNotEmpty{$\mathcal{S}$}} {
        $n_0$ $\leftarrow$ $n_0\times2$\;
        $s$ $\leftarrow$ $s/2$\;
        \If{\pop{$\mathcal{S}$}}{
            \emph{Reattach the last row of each sub-system (See Algorithm \ref{alg:reattach})}\;
            $n_a$ $\leftarrow$ $n_a + s$\;
            $n_0$ $\leftarrow$ $n_0 + 1$\;
        }
    }
    \tag $\leftarrow$ \sendData{$\sdata=\widetilde{\bm x}_i$, $\drank = (p+i-1)\mod p$}\;
    
    $\widetilde{\bm x}_{i+1}$ $\leftarrow$ \getData{\tag, $\srank = (i+1)\mod p$}\;
    
    ${\bm x}_i$ $\leftarrow$ ${\bm x}_i - {\bm Y}_i\widetilde{\bm x}_i - {\bm Z}_i\widetilde{\bm x}_{i+1}$\;
    
    \caption{Implementation of in-place solver. The rank of partition is zero-based.}
    \label{alg:main}
\end{algorithm}

\begin{algorithm}[htbp]
    \SetKwData{sdata}{send\_buffer}
    \SetKwData{drank}{dest\_rank}
    \SetKwData{srank}{src\_rank}
    \SetKwData{tag}{tag}
    \SetKwData{tagA}{tag\_a}
    \SetKwData{tagB}{tag\_b}
    \SetKwFunction{genPCR}{generalizedPCR}
    \SetKwFunction{getData}{getFromPartition}
    \SetKwFunction{sendData}{sendToPartition}
    \SetKwFunction{emptyStack}{initEmptyStack}
    \SetKwFunction{push}{stackPush}
    \SetKwFunction{pop}{stackPop}
    \SetKwInOut{Input}{in}
    \SetKwInOut{Output}{in/out}
    
    \If{$n_a\le i < (n_a+s)$}{
        \tagA $\leftarrow$ \sendData{$\sdata=\{\widehat{\bm L}_i, {\bm W}_i, \widehat{\bm U}_i, \widetilde{\bm x}_i\}$, $\drank = i-s$}\;
        \tagB $\leftarrow$ \sendData{$\sdata=\{\widehat{\bm L}_i, {\bm W}_i, \widehat{\bm U}_i, \widetilde{\bm x}_i\}$, $\drank = i-n_a$}\;
    }
    \If{$n_a\le (i+s) < (n_a+s)$}{
        $\{\widehat{\bm L}_{i+s}, {\bm W}_{i+s}, \widehat{\bm U}_{i+s}, \widetilde{\bm x}_{i+s}\}$ $\leftarrow$ \getData{\tagA, $\srank = i+s$}\;
        ${\bm W}_i$ $\leftarrow$ ${\bm W}_i - \widehat{\bm U}_i{\bm W}_{i+s}^{-1}\widehat{\bm L}_{i+s}$\;
        $\widetilde{\bm x}_i$ $\leftarrow$ $\widetilde{\bm x}_i - \widehat{\bm U}_i{\bm W}_{i+s}^{-1}\widetilde{\bm x}_{i+s}$\;
        $\widehat{\bm U}_i$ $\leftarrow$ $-\widehat{\bm U}_i{\bm W}_{i+s}^{-1}\widehat{\bm U}_{i+s}$\;
    }
    \If{$n_a\le (i+n_a) < (n_a+s)$}{
        $\{\widehat{\bm L}_{i-s}, {\bm W}_{i-s}, \widehat{\bm U}_{i-s}, \widetilde{\bm x}_{i-s}\}$ $\leftarrow$ \getData{\tagB, $\srank = i+n_a$}\;
        ${\bm W}_i$ $\leftarrow$ ${\bm W}_i - \widehat{\bm L}_i{\bm W}_{i-s}^{-1}\widehat{\bm U}_{i-s}$\;
        $\widetilde{\bm x}_i$ $\leftarrow$ $\widetilde{\bm x}_i - \widehat{\bm L}_i{\bm W}_{i-s}^{-1}\widetilde{\bm x}_{i-s}$\;
        $\widehat{\bm L}_i$ $\leftarrow$ $-\widehat{\bm L}_i{\bm W}_{i-s}^{-1}\widehat{\bm L}_{i-s}$\;
    }
    
    \caption{Detaching process in Algorithm \ref{alg:main}.}
    \label{alg:detach}
\end{algorithm}
\begin{algorithm}[htbp]
    \SetKwData{sdata}{send\_buffer}
    \SetKwData{drank}{dest\_rank}
    \SetKwData{srank}{src\_rank}
    \SetKwData{tag}{tag}
    \SetKwData{tagA}{tag\_a}
    \SetKwData{tagB}{tag\_b}
    \SetKwFunction{genPCR}{generalizedPCR}
    \SetKwFunction{getData}{getFromPartition}
    \SetKwFunction{sendData}{sendToPartition}
    \SetKwFunction{emptyStack}{initEmptyStack}
    \SetKwFunction{push}{stackPush}
    \SetKwFunction{pop}{stackPop}
    \SetKwInOut{Input}{in}
    \SetKwInOut{Output}{in/out}
    
    \If{$i < n_a$}{
        \tagA $\leftarrow$ \sendData{$\sdata=\{\widehat{\bm L}_i, {\bm W}_i, \widehat{\bm U}_i, \widetilde{\bm x}_i\}$, $\drank = (n_a+i-s)\mod n_a$}\;
        \tagB $\leftarrow$ \sendData{$\sdata=\{\widehat{\bm L}_i, {\bm W}_i, \widehat{\bm U}_i, \widetilde{\bm x}_i\}$, $\drank = (i+s)\mod n_a$}\;
        $\{\widehat{\bm L}_{i+s}, {\bm W}_{i+s}, \widehat{\bm U}_{i+s}, \widetilde{\bm x}_{i+s}\}$ $\leftarrow$ \getData{\tagA, $\srank = (i+s)\mod n_a$}\;
        $\{\widehat{\bm L}_{i-s}, {\bm W}_{i-s}, \widehat{\bm U}_{i-s}, \widetilde{\bm x}_{i-s}\}$ $\leftarrow$ \getData{\tagB, $\srank = (n_a+i-s)\mod n_a$}\;
        ${\bm W}_i$ $\leftarrow$ ${\bm W}_i - \widehat{\bm U}_i{\bm W}^{-1}_{i+s}\widehat{\bm L}_{i+s} - \widehat{\bm L}_i{\bm W}^{-1}_{i-s}\widehat{\bm U}_{i-s}$\;
        $\widetilde{\bm x}_i$ $\leftarrow$ $\widetilde{\bm x}_i - \widehat{\bm U}_i{\bm W}^{-1}_{i+s}\widetilde{\bm x}_{x+s} - \widehat{\bm L}_i{\bm W}^{-1}_{i-s}\widetilde{\bm x}_{i-s}$\;
        $\widehat{\bm L}_i$ $\leftarrow$ $-\widehat{\bm L}_i{\bm W}^{-1}_{i-s}\widehat{\bm L}_{i-s}$\;
        $\widehat{\bm U}_i$ $\leftarrow$ $-\widehat{\bm U}_i{\bm W}^{-1}_{i+s}\widehat{\bm U}_{i+s}$\;
    }
    
    \caption{Block PCR process in Algorithm \ref{alg:main}.}
    \label{alg:blockPCR}
\end{algorithm}
\begin{algorithm}[htbp]
    \SetKwData{sdata}{send\_buffer}
    \SetKwData{drank}{dest\_rank}
    \SetKwData{srank}{src\_rank}
    \SetKwData{tag}{tag}
    \SetKwData{tagA}{tag\_a}
    \SetKwData{tagB}{tag\_b}
    \SetKwFunction{genPCR}{generalizedPCR}
    \SetKwFunction{getData}{getFromPartition}
    \SetKwFunction{sendData}{sendToPartition}
    \SetKwFunction{emptyStack}{initEmptyStack}
    \SetKwFunction{push}{stackPush}
    \SetKwFunction{pop}{stackPop}
    \SetKwInOut{Input}{in}
    \SetKwInOut{Output}{in/out}
    
    \If{$n_a\le (i+s) < (n_a+s)$}{
        \tagB $\leftarrow$ \sendData{$\sdata=\widetilde{\bm x}_i$, $\drank = i+s$}\;
    }
    \If{$n_a\le (i+n_a) < (n_a+s)$}{
        \tagA $\leftarrow$ \sendData{$\sdata=\widetilde{\bm x}_i$, $\drank = i+n_a$}\;
    }
    \If{$n_a\le i < (n_a+s)$}{
        $\widetilde{\bm x}_{i+s}$ $\leftarrow$ \getData{\tagA, $\srank = i-n_a$}\;
        $\widetilde{\bm x}_{i-s}$ $\leftarrow$ \getData{\tagB, $\srank = i-s$}\;
        $\widetilde{\bm x}_i$ $\leftarrow$ $\widetilde{\bm x}_i - \widehat{\bm L}_i\widetilde{\bm x}_{i-s} - \widehat{\bm U}_i\widetilde{\bm x}_{i+s}$\;
        $\widetilde{\bm x}_i$ $\leftarrow$ ${\bm W}_i^{-1}\widetilde{\bm x}_i$\;
    }
    
    \caption{Reattaching process in Algorithm \ref{alg:main}.}
    \label{alg:reattach}
\end{algorithm}

\begin{figure}
    \centering
    \subfloat[]{\scalebox{0.4}{\tikzset{
    thick/.style={line width=2.5pt},
    thin/.style={line width=1.0pt},
    block l/.pic={
        \fill[magenta!50!white] (0,0) -- (2,-2) |- cycle;
        \draw[draw=black, thick] (-6,0) rectangle (2,-2);
        \draw[draw=black, thin, dashed] (0,0) -- (0,-2);
        \draw[thin] ( 2, 0.2) --++ (0, 1.8);
        \draw[thin] ( 0, 0.2) --++ (0, 0.8);
        \draw[thin] (-6, 0.2) --++ (0, 1.8);
        \draw[thin, <->, >=stealth] (2,0.6) --++ (-2, 0);
        \draw[thin, <->, >=stealth] (2,1.6) --++ (-8, 0);
        \node [anchor=center] at ( 1, 0.8) {\LARGE $r$};
        \node [anchor=center] at (-2, 2.0) {\LARGE $N_{i-1}$};
    },
    block y/.pic={
        \fill[cyan!20!white] (0, 6) rectangle (2, 0);
        \fill[cyan!60!white] (0, 0) rectangle (2,-2);
        \draw[thin, dashed] (0, 0) -- (2, 0);
        \draw[draw=black, thick] (0,6) rectangle (2,-2);
        \draw[thin] (2.2, 6) --++ (1.8,0);
        \draw[thin] (2.2, 0) --++ (0.8,0);
        \draw[thin] (2.2,-2) --++ (1.8,0);
        \draw[thin, <->, >=stealth] (2.6,-2) --++ (0,2);
        \draw[thin, <->, >=stealth] (3.8,-2) --++ (0,8);
        \node[anchor=west] at (2.8,-1) {\LARGE $r$};
        \node[anchor=west] at (4.0, 2) {\LARGE $N_{i-1}$};
    },
}

\begin{tikzpicture}
    \pic at (0,0) {block l};
    \pic at (4,0) {block y};
    \node [anchor=center] at (-2, -3) {\LARGE $\widetilde{\bm L}_i$};
    \node [anchor=center] at ( 5, -3) {\LARGE ${\bm S}_{i-1}$, ${\bm R}_{i-1}$, or ${\bm y}_{i-1}$};
\end{tikzpicture}}\label{fig:comm_pattern}}
    \qquad
    \subfloat[]{\scalebox{0.4}{\tikzset{
    thick/.style={line width=2.5pt},
    thin/.style={line width=1.0pt},
    block l/.pic={
        \fill[magenta!50!white] (-6,0) -- (-4,-2) -| cycle;
        \draw[draw=black, thick] (-6,0) rectangle (2,-2);
        \draw[draw=black, thin, dashed] (-4,0) -- (-4,-2);
        \draw[thin] ( 2, 0.2) --++ (0, 1.8);
        \draw[thin] (-4, 0.2) --++ (0, 0.8);
        \draw[thin] (-6, 0.2) --++ (0, 1.8);
        \draw[thin, <->, >=stealth] (-4,0.6) --++ (-2, 0);
        \draw[thin, <->, >=stealth] (2,1.6) --++ (-8, 0);
        \node [anchor=center] at (-5, 0.8) {\LARGE $r$};
        \node [anchor=center] at (-2, 2.0) {\LARGE $N_i$};
    },
    block y/.pic={
        \fill[cyan!20!white] (0, 4) rectangle (2,-2);
        \fill[cyan!60!white] (0, 6) rectangle (2, 4);
        \draw[thin, dashed] (0, 4) --++ (2, 0);
        \draw[draw=black, thick] (0,6) rectangle (2,-2);
        \draw[thin] (2.2, 6) --++ (1.8,0);
        \draw[thin] (2.2, 4) --++ (0.8,0);
        \draw[thin] (2.2,-2) --++ (1.8,0);
        \draw[thin, <->, >=stealth] (2.6, 4) --++ (0,2);
        \draw[thin, <->, >=stealth] (3.8,-2) --++ (0,8);
        \node[anchor=west] at (2.8, 5) {\LARGE $r$};
        \node[anchor=west] at (4.0, 2) {\LARGE $N_i$};
    },
}

\begin{tikzpicture}
    \pic at (0,0) {block l};
    \pic at (4,0) {block y};
    \node [anchor=center] at (-2, -3) {\LARGE $\widetilde{\bm U}_i$};
    \node [anchor=center] at ( 5, -3) {\LARGE ${\bm S}_{i}$, ${\bm R}_{i}$, or ${\bm y}_{i}$};
\end{tikzpicture}}\label{fig:mult_pattern}}
    \caption{Communication and multiplication patterns: \protect\subref{fig:comm_pattern} $\widetilde{\bm L}_i{\bm S}_{i-1}$, $\widetilde{\bm L}_i{\bm R}_{i-1}$, and $\widetilde{\bm L}_i{\bm y}_{i-1}$; \protect\subref{fig:mult_pattern} $\widetilde{\bm U}_i{\bm S}_{i}$, $\widetilde{\bm U}_i{\bm R}_{i}$, and $\widetilde{\bm U}_i{\bm y}_{i}$}
    \label{fig:comm_mult_pattern}
\end{figure}
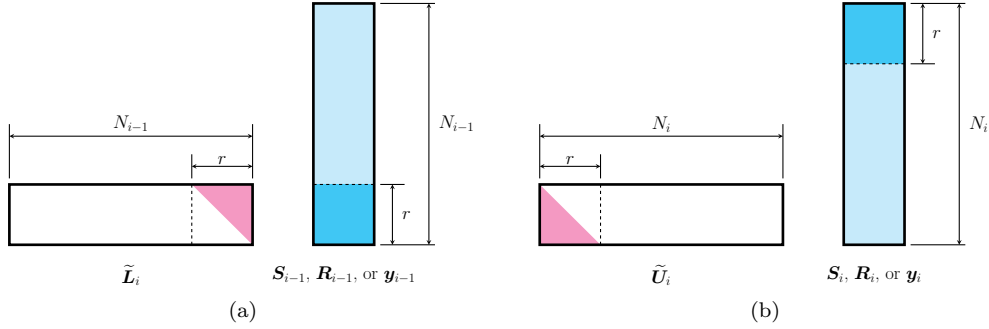

Throughout the solution process, the terms $\widetilde{\bm L}_i{\bm S}_{i-1}$, $\widetilde{\bm L}_i{\bm R}_{i-1}$, and $\widetilde{\bm L}_i{\bm y}_{i-1}$ require data transfer from partition $i-1$ to $i$, and the term ${\bm R}_i\widetilde{\bm x}_{i+1}$ implies the data transfer from partition $i+1$ to $i$.
It is important to emphasize that the sparsity pattern of the matrix $\widetilde{\bm{L}}_i$ results in only a fraction of the allocated data in ${\bm y}_{i-1}$, ${\bm R}_{i-1}$ and ${\bm S}_{i-1}$ exchanged across neighbor data partitions as illustrated in Figure \ref{fig:BLU1} and Figure \ref{fig:comm_pattern}.
For the banded matrix, $\bm{A}$ with a bandwidth is $w = 2r+1$, only the last $r$ columns in $\widetilde{\bm L}_i$ are non-trivial. Therefore, only the last $r$ columns in ${\bm S}_{i-1}$, ${\bm R}_{i-1}$, and ${\bm y}_{i-1}$ are needed for neighbor communication. Similarly, the matrix products involving $\bm{U}_{i}$ can be computing very efficiently due to its sparsity pattern as shown in Figure \ref{fig:comm_mult_pattern}.
If the number of rows in each partition is much larger than the system bandwidth ($N_i \gg r$), significant reduction of data size for communication and multiplication can be achieved. The reduced system $\widehat{\bm A}\widetilde{\bm x} = \widehat{\bm b}$ (Equation \ref{eqn:sub-system}) is solved on  distributed memory, and each parallel reduction step requires data communication between neighboring partitions. If ${\bm A}$ is acyclic, then $\widehat{\bm A}$ can be solved with the classic block PCR, although the proposed algorithm can still be used by setting the cyclic entries to zero, and the number of the parallel reduction steps is $\lceil{\log_2p}\rceil$, where $p$ is the number of partitions. If ${\bm A}$ is cyclic, using the proposed algorithm, the number of the parallel reduction steps is $\lfloor{\log_2p}\rfloor$. In addition, if $p$ is not a power of $2$, the number of rows that are involved in the detaching and reattaching throughout the solving process equals $p - 2^{\lfloor{\log_2p}\rfloor}$, and the numbers of the parallel detaching and reattaching steps are $\left\{\sum_{n=0}^{ \lfloor{\log_2p}\rfloor}\left(\lfloor{2^{-n}p}\rfloor \mod 2\right)\right\}-1$.

In the motivating applications, such as evaluating derivatives using compact finite differences in a multiphysics application, ${\bm A}{\bm x} = {\bm b}$ is frequently solved with varying ${\bm b}$ but constant ${\bm A}$. Noticing that the construction of $\widehat{\bm L}_i$, $\widehat{\bm D}_i$, $\widehat{\bm U}_i$, and ${\bm D}_i$, does not require the right-hand-side, $\bm b$, such construction is needed only once, and the original matrix can be pre-factorized. During the pre-factorization, the reduction coefficients on each stage $k_{+j}$ and $k_{-j}$, and the information needed to solve $\widehat{\bm A}\widetilde{\bm x}=\widehat{\bm b}$, can be calculated and stored. During the solution process, Equation (\ref{eqn:yi}) and Equation (\ref{eqn:bi_hat}) are needed to construct the right-hand-side of the reduced system to solve $\widetilde{\bm x}_i$. Finally, Equation (\ref{eqn:xi_app}) used to solve for ${\bm x}_i$.

\section{Performance}
\label{sec:performance}
In this section, the performance of the linear solver is demonstrated both in isolation and in the context of a representative fluid mechanics application problem. All tests in this section were performed on the \textit{Summit} supercomputer at the Oak Ridge Leadership Computing Facility (OLCF) at Oak Ridge National Laboratory (ORNL) \cite{summit_2018}. Each \textit{Summit} node consists of 6 NVIDIA Tesla V100 GPUs and 2 IBM Power 9 processors. The nodes on the system are connected with Mellanox EDR 100G Infiniband interconnect, arranged in a non-blocking fat tree topology.

\begin{figure}[htbp]
    \centering
    \includegraphics[width=0.4\textwidth]{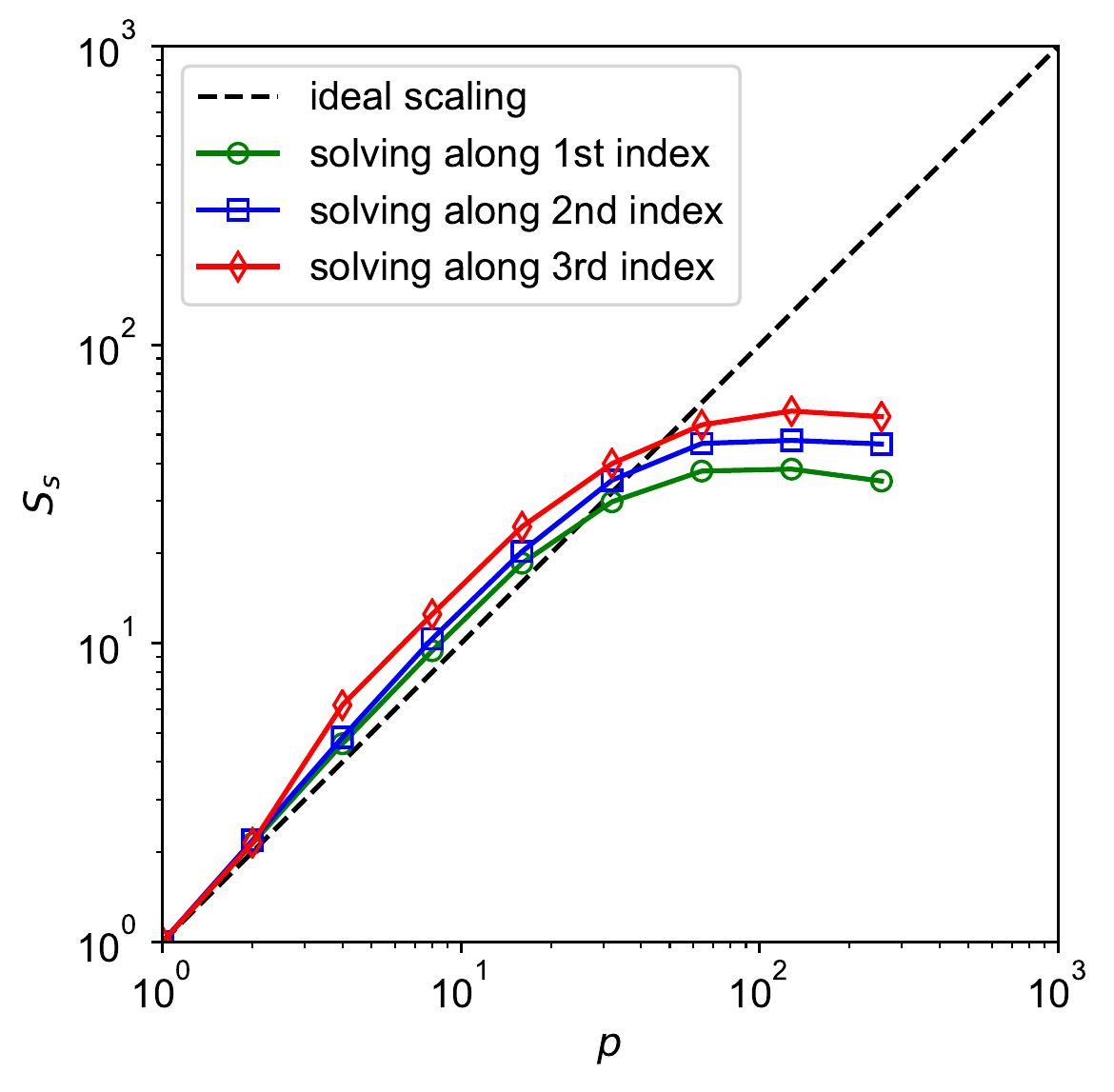}
    \caption{Measured strong scaling of the linear solver for each of the coordinate indices. The curve for each index is normalized by its own single-GPU time, so all speedups start at unity.}
    \label{fig:linsol_strong_scaling}
\end{figure}

For the linear solver alone, both strong and weak scaling results are presented for solving $\bm A \bm x = \bm b$, where $\bm A$ is a cyclic tridiagonal system with bands given by $\bm A = \mathcal{B}~[~1/3,~1,~1/3~]$. This linear system represents the left hand side of the sixth order compact first derivative scheme on a periodic domain \cite{lele1992compact}. For all linear solver scaling tests, the linear system is solved $1000$ times, and speedup based on the average time is reported. In the strong scaling test, the dimension of ${\bm A}$ is $8192 \times 8192$, and the linear system is solved $256^2$ times in parallel, i.e., the dimensions of ${\bm b}$ and ${\bm x}$ are $8192\times256^2$. In the context of the compact finite difference scheme, this is equivalent to computing a spatial derivative along a column of 3D Cartesian grid partitions, where the grid dimension is $8192$ along the solving direction and $256\times256$ perpendicular to the solving direction. For example, when solving along the first index, the grid is $8192 \times256\times256$ mesh. As the number of GPUs used is increased, the domain is decomposed equally along the solving direction so that each partition has the size of $(8192/p)\times256\times256$. When solving along other directions, the dimensions are permuted correspondingly. Some small differences in performance among the directions are expected because of memory striding. In this implementation, right memory layout is used, where the third index maps to contiguous memory. The strong scaling speedup, $S_s$, is defined as
\begin{equation}
    S_s (p) = \frac{T_1}{T_{p}}
\end{equation}
where $T_p$ is the wall time when using $p$ GPUs. The strong scaling results for each index direction are shown in Figure \ref{fig:linsol_strong_scaling}.

The strong scaling behavior can be understood by considering the different ways parallelism is expressed in the algorithm. First, there is data parallelism along the non-solving directions. Second, there is parallelism along the solving direction within each PCR step on shared memory. Finally, there is parallelism to solve the reduced system along the solving direction among the different GPUs, and PCR on distributed memory is carried out at this level. When $p$ is small, PCR within each shared memory dominates the computational cost. Due to the large workload on each GPU, the parallelism of the elementwise operations as well as each PCR step on shared memory is not fully parallelized. The floating point operations of the first kind of parallelism scales with $\mathcal{O}(N_i)$, and the number of steps of the second kind of parallelism scales with $\mathcal{O}(N_i\log_2 N_i)$. Therefore, as $p$ increases, both of the aforementioned types of parallelism are exploited, which makes the scaling superlinear initially. However, as $p$ keeps increasing and the workload per GPU decreases more, the cost of PCR on the distributed memory dominates, and communication becomes the time limiting factor. Then speedup reaches a plateau, as seen in Figure \ref{fig:linsol_strong_scaling}. As $p$ increases further, the number of communication stages, which increases with $\log_2 p$, becomes non-negligible, the speedup will decline slowly. At $p=32$, approximately where the three curves cross the ideal scaling line, the data chunk on each GPU is $256\times256\times256$. This is the chunk size used as the basis of the weak scaling tests, which are discussed next.

\begin{figure}[htbp]
    \centering
    \subfloat[]
    {\includegraphics[width=0.4\textwidth]{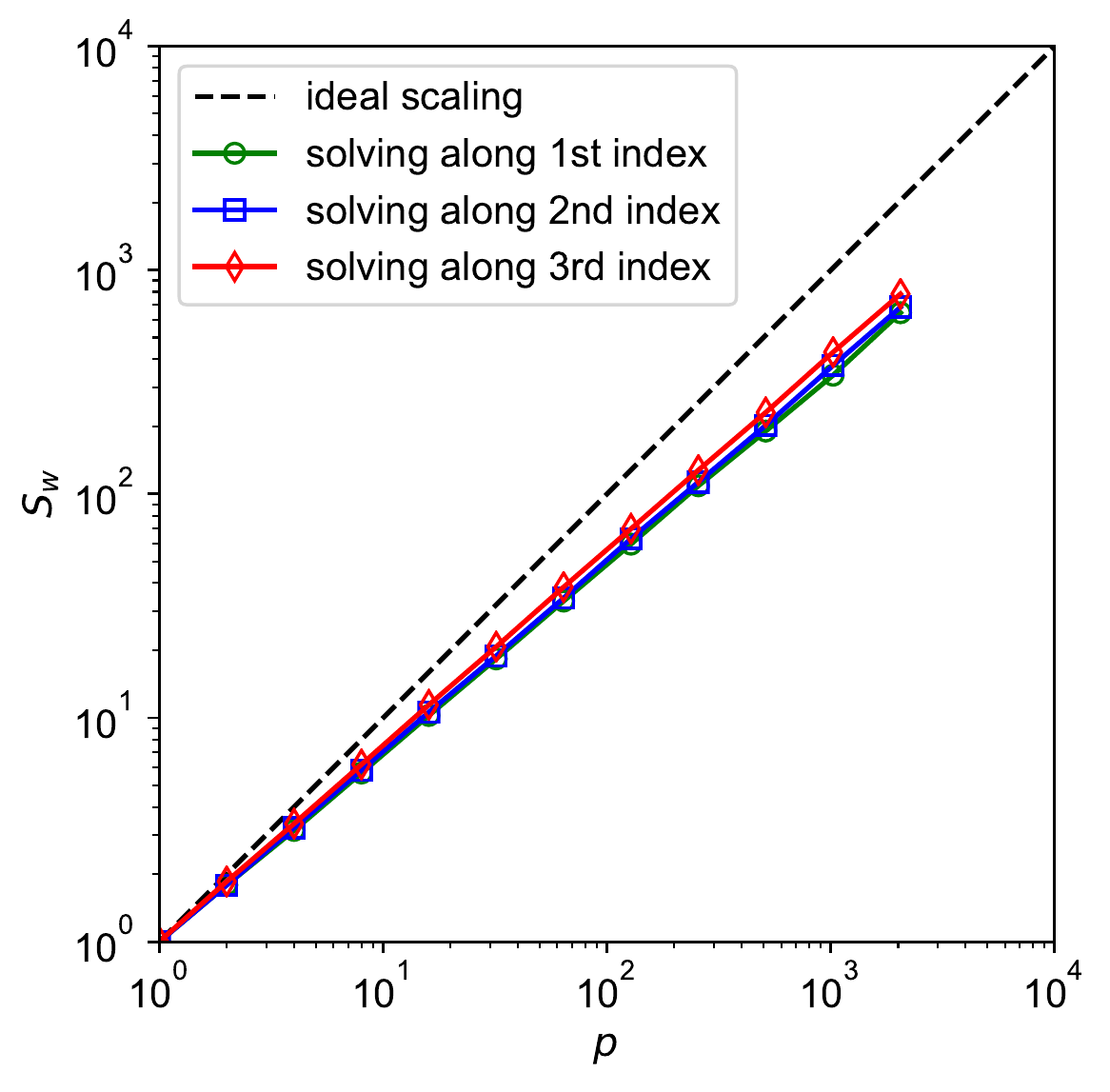}\label{fig:linsol_scaling_exp}}
    \qquad
    \subfloat[]
    {\includegraphics[width=0.4\textwidth]{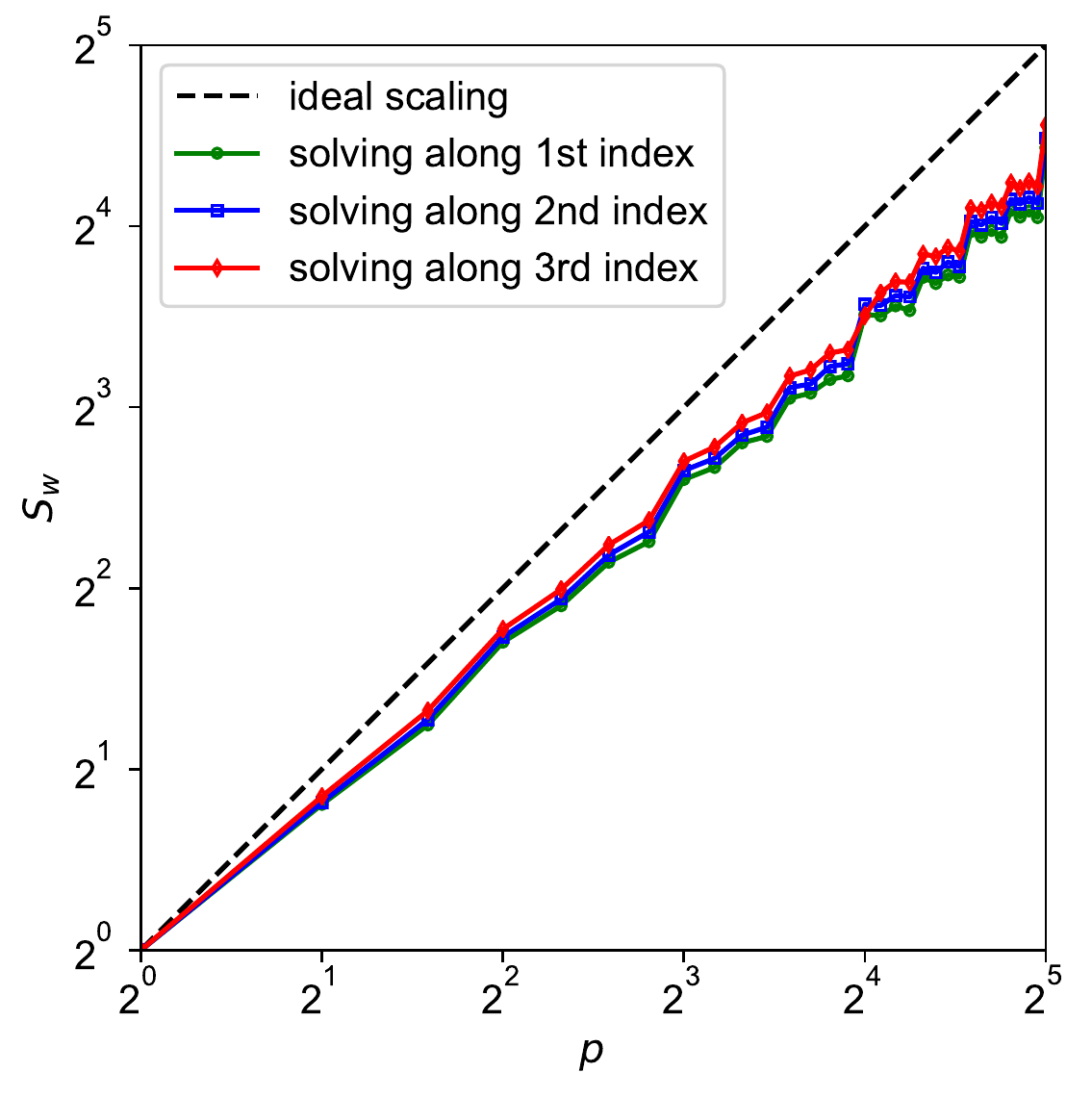}\label{fig:linsol_scaling_lin}}
    \caption{Measured weak scaling of the linear solver: \protect\subref{fig:linsol_scaling_exp} number of GPUs increasing in powers of $2$; \protect\subref{fig:linsol_scaling_lin} number of GPUs increasing linearly. Data is stored on the left memory layout where the 3rd index is the fast looping index.}
    \label{fig:linsol_scaling}
\end{figure}


The weak scaling performance is shown for solving $\bm A \bm x = \bm b$, with the same matrix $\bm A$ as in the strong scaling test. The computational domain is partitioned along the solve direction into cubic sub-domains of size $N_0 \times N_0 \times N_0$, so $\bm A$ is $pN_0\times pN_0$, and it is solved $N_0^2$ times in parallel. For all the weak scaling tests, $N_0=256$ is chosen, so that each GPU operates on a chunk of data that is $256^3$. The weak scaling results for the isolated linear system solve are presented in two ways in Figure \ref{fig:linsol_scaling}. Here, the weak scaling ``speedup'', $S_w$, is reported:
\begin{equation}
    S_w (p) = \frac{p \times T_1}{T_p}
\end{equation}
First, in Figure \ref{fig:linsol_scaling_exp}, the number of GPUs used is always a power of 2, from 1 to 2048. As the number of GPUs is increased in the weak scaling test, the dimension of only one coordinate direction is increased at a time, and all three directions are tested.  The grid sizes for the series of tests for the first index, for example, are $[256\times 256\times256]$, ~$[512\times 256\times256]$, ~$[1024\times 256\times256]$, ~etc. Second, in Figure \ref{fig:linsol_scaling_lin}, the number of GPUs used increases linearly from 1 to 32, to show the effect of a non-ideal problem decomposition on performance. Also, the differences in scaling among the index directions are very small, meaning that in a large scale 3D problem, no one direction will dominate the computational cost. These results show that the scaling of the linear solver is reasonably good up to a very large number of GPUs. For context, the last data point comes from running on $2048$ GPUs on \textit{Summit}, or about $8\%$ of the entire machine. This test exercises one coordinate direction at a time on a column of the domain decomposition, comparable to the highlighted partitions in Figure \ref{fig:grid-partition} in order to predict the performance in a realistic computation application. Accordingly, the last data point represents the intended 3D equal size domain decomposition used by the linear solver in a production size simulation which uses $2048^3$ GPUs.

\begin{figure}[htbp]
    \centering
    \includegraphics[width=0.4\textwidth]{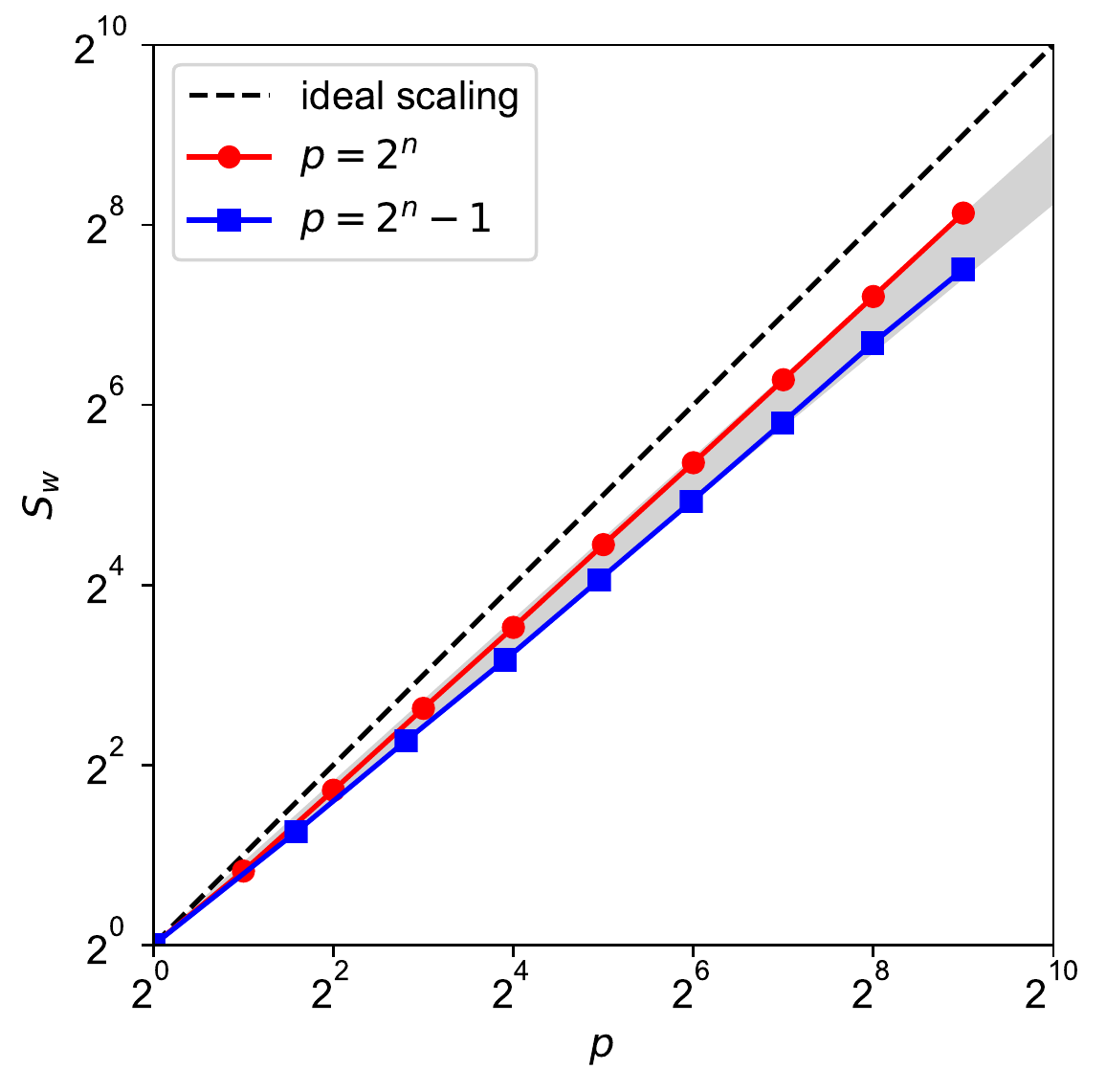}
    \caption{Measured weak scaling of the linear solver using best-case ($2^n$) and worst-case ($2^n-1$) numbers of GPUs, solving along the 1st index.}
    \label{fig:linsol_scaling_nonideal}
\end{figure}

These results also show that while the solver scales best when using a number of GPUs equal to a power of 2, its performance is degraded using odd or even prime numbers of GPUs. This occurs because additional work in the form of detach-reattach steps is required when not using power of 2 number of GPUs. The worst case scenario in terms of additional work required is to use a number of GPUs equal to $2^n-1$. This choice requires $p-1$ stages of PCR and $p-1$ stages of detach-reattach operations. An additional series of weak scaling results is presented in Figure \ref{fig:linsol_scaling_nonideal}, which compares the weak scaling performance of using $2^n$ vs. $2^n-1$ GPUs. Only the 1st index direction is shown, since the results are qualitatively the same for all directions.  Depending on the specific machine and application size, it may not always be practical to use a number of GPUs that is a power of 2. As a result, it is expected that the practical weak scaling behavior of this algorithm lies in the range between the curves in Figure \ref{fig:linsol_scaling_nonideal}. Both curves are demonstrated to be linear over the range tested. This is expected based on how the number of PCR steps and attach-reattach operations scales with the number of processes. Since the lines have different slopes, this means that the relative benefit of using the ideal number of GPUs becomes greater as the problem size is increased.

\begin{figure}[htbp]
    \centering
    \includegraphics[width=0.4\textwidth]{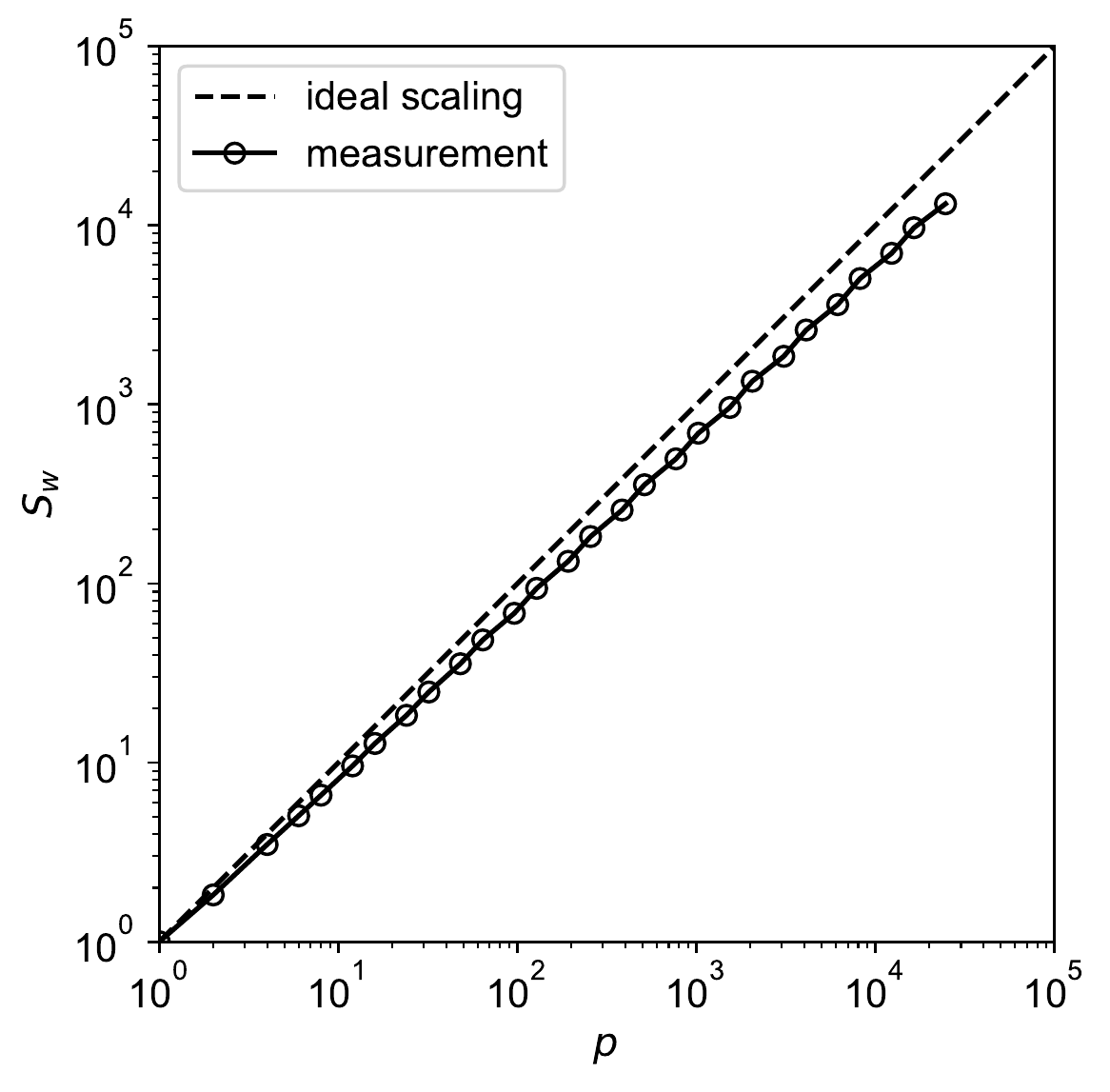}
    \caption{Measured weak scaling ($256^3$ grid point per GPU) of a Navier-Stokes solver on the Taylor-Green vortex problem using compact finite difference and interpolation methods. Data is reported using both $4 \times 2^n$ and $6\times2^n$ GPUs per node on \textit{Summit}.}
    \label{fig:TGV_scaling}
\end{figure}

Finally, weak scaling is demonstrated on a fluid mechanics application -- the direct numerical simulation of a Taylor-Green vortex problem at the Reynolds number of $1600$ and Mach number of $0.08$ \cite{bull2015simulation} -- by using a compressible Navier-Stokes direct numerical simulation solver. The simulations were conducted using the sixth-order staggered compact finite difference schemes and compact interpolators for spatial discretization \cite{nagarajan2003robust}. The details of the problem description and numerical formulation are illustrated in \ref{sec:tgv}. For each weak scaling test, a constant time step determined by stability requirements was used, and wall time data was collected for 100 time steps. The computational cost is dominated by calculating derivatives and interpolations in the Navier-Stokes equations, which involves solving linear systems similar to the one above.
The solution at a representative time is visualized in Figure \ref{fig:TGV_qcriterion}.  This test is useful because its domain and domain decomposition are much more realistic than the isolated linear solver test, and because it involves approximately equal numbers of linear solves along all three coordinate indices. Finally, it tests whether the linear solver performance enables good scaling on a practical problem. Like the first linear solver test, scaling is reported in powers of 2, but quantities of GPUs of $6\times2^n$ were also tested. This second series corresponds to full utilization of \textit{Summit} nodes, which have 6 GPUs each. The weak scaling results, including both setups, are shown in Figure \ref{fig:TGV_scaling}, which demonstrates excellent scaling up to 24576 GPUs, or 89\% of the nodes on \textit{Summit}. 

The reason that Taylor-Green vortex problem scales better than the linear solver test is due to the 3D domain decomposition. The linear solver weak scaling tested an extreme scenario, with a 1D domain decomposition. This would only be appropriate for a domain with one dimension much longer than the other two. Such an aspect ratio is not typical for simulations of turbulent flow.

\section{Conclusions}

In this work, a direct linear solver for compact banded systems is presented and demonstrated to have scalable performance on a petascale GPU platform. The algorithm is applicable for a wide variety of high performance computing platforms with heterogeneous computing capabilities. The sparsity patterns that result in the factorized matrix blocks are leveraged in the overall algorithm to avoid large data transfers across the distributed memory partitions and to reduce the floating point operational cost of matrix-matrix multiplications. As such, the proposed algorithm has significant advantages over conventional strategies that involve ``all-to-all'' communication patterns. These advantages thereby enable the proposed algorithm to be suitable for distributed heterogeneous computing environments requiring programming paradigms such as ``MPI+X'', and to reduce the strong performance dependence on the underlying network topology. The weak scalability is shown on a canonical 3D periodic Navier-Stokes problem using compact finite difference and interpolation schemes involving cyclic banded tridiagonal linear systems. The algorithm works on a flexible number of distributed memory partitions and optimal performance is recovered when the number of ranks is a power-of-two. This work is directly beneficial to the large scale computations of a wide range of partial differential equation problems using compact numerical schemes such as in fluid mechanics, solid mechanics, and electromagnetics.

\section*{Acknowledgements}
The authors are grateful to Professor Eric Darve for helpful comments.
This research used resources of the Oak Ridge Leadership Computing Facility at the Oak Ridge National Laboratory, which is supported by the Office of Science of the U.S. Department of Energy under Contract No. DE-AC05-00OR22725 \cite{summit_2018}. 
This work also used the Extreme Science and Engineering Discovery Environment (XSEDE), which is supported by National Science Foundation grant number ACI-1548562 \cite{xsede}. This work used XSEDE resources \textit{Bridges} and \textit{Comet} through allocation TG-CCR130001. 

\appendix
\section{Taylor-Green vortex}\label{sec:tgv}
\subsection{Problem description}
The Taylor-Green vortex problem is a well-established fluid mechanics problem defined on 3D periodic domain, ${\bm x} \in [0, 2\pi l)\times[0, 2\pi l)\times[0, 2\pi l)$, where $l$ is a characteristic length. The tests used in this work were conducted by solving the compressible Navier-Stokes equations.
\begin{equation}
    \frac{\partial{\bm \phi}}{\partial t} + \nabla\cdot{\bm F} + \nabla\cdot{\bm G} = 0
\end{equation}
where ${\bm \phi}$ is the set of the conservative variables; ${\bm F}$ is the set of inviscid fluxes; and ${\bm G}$ is the set of diffusive fluxes. They are defined as
\begin{align}
    {\bm \phi} &=
    \begin{bmatrix}
        \rho \\ \rho {\bm u} \\ \rho (e + {\bm u}\cdot{\bm u}/2)
    \end{bmatrix}
    \\
    {\bm F} &=
    \begin{bmatrix}
        \rho {\bm u} \\
        \rho \left({\bm u}\otimes{\bm u}\right) + P{\bm I} \\
        {\bm u}\left(\rho e + \rho{\bm u}\cdot{\bm u}/2 + P\right)
    \end{bmatrix}
    \\
    {\bm G} &=
    \begin{bmatrix}
        0 \\
        -{\bm \sigma} \\
        {\bm q} - {\bm u}\cdot{\bm\sigma}
    \end{bmatrix}
\end{align}
where $\rho$ is the density; ${\bm u} = [u, v, w]^T$ is the velocity vector; $P$ is the pressure; ${\bm I}$ is the identity tensor; $e$ is the specific internal energy, ${\bm \sigma}$ is the viscous stress tensor; and ${\bm q}$ is the heat flux. The fluid is treated as ideal gas with the following equation of state.
\begin{equation}
    P = \rho R T
\end{equation}
where $R$ is the specific gas constant; and $T$ is the temperature. Accordingly, the internal energy is
\begin{equation}
    e = \frac{RT}{\gamma - 1}
\end{equation}
where $\gamma$ is the ratio of specific heat. The viscous stress tensor is modeled as
\begin{equation}
    {\bm\sigma} = \mu\left[(\nabla{\bm u}) + (\nabla{\bm u})^T\right]+ \left(\beta - \frac{2}{3}\mu\right)\left(\nabla\cdot{\bm u}\right){\bm I}
\end{equation}
where $\mu$ is the dynamic shear viscosity; and $\beta$ is the bulk viscosity. For the simulations used in this work, $\beta = 0$ and $\mu$ is set to be a constant determined from the Reynolds number, $\mathrm{Re}$.
\begin{equation}
    \mathrm{Re} = \frac{\rho_0 Vl}{\mu}
\end{equation}
where $\rho_0$ is the mean density as well as the initial density of the fluid, and $V$ is a characteristic velocity. The heat flux $\bm q$ is computed based on the Fourier law
\begin{equation}
    {\bm q} = -\kappa\nabla T
\end{equation}
where $\kappa$ is the heat conductivity controlled by the Prandtl number, $\mathrm{Pr}$, defined in the following.
\begin{equation}
    \mathrm{Pr} = \frac{\gamma R\mu}{(\gamma-1)\kappa}
\end{equation}
The initial velocity, $[u_0, v_0, w_0]^T$, and pressure, $P_0$, fields are set as \cite{bull2015simulation}
\begin{align}
    u_0 &= V\sin(x/l)\cos(y/l)\cos(z/l)
    \\
    v_0 &= -V\cos(x/l)\sin(y/l)\cos(z/l)
    \\
    w_0 &= 0
    \\
    P_0 &= P_\mathrm{ref} + \frac{\rho_0V^2}{16}\left[\cos(2x/l) + \cos(2y/l)\right]\left[\cos(2z/l) + 2\right]
\end{align}
where $l = 1$, $V = 1$, $\rho_0 = 1$, and $P_\mathrm{ref} = 100$. The Reynolds number and Prandtl number are set to $\mathrm{Re}=1600$ and $\mathrm{Pr}=0.7$, respectively. The specific gas constant is set to unity, and the specific heat ratio $\gamma = 5/3$, so that the Mach number, $\mathrm{Ma}$, consistent with the initial condition, is approximately $0.08$, which is calculated in the following based on its definition.
\begin{equation}
    \mathrm{Ma} = \frac{V}{\sqrt{\gamma P_\mathrm{ref} / \rho_0}}
\end{equation}

\begin{figure}[htbp]
    \centering
    \subfloat[]
    {\includegraphics[width=0.35\textwidth]{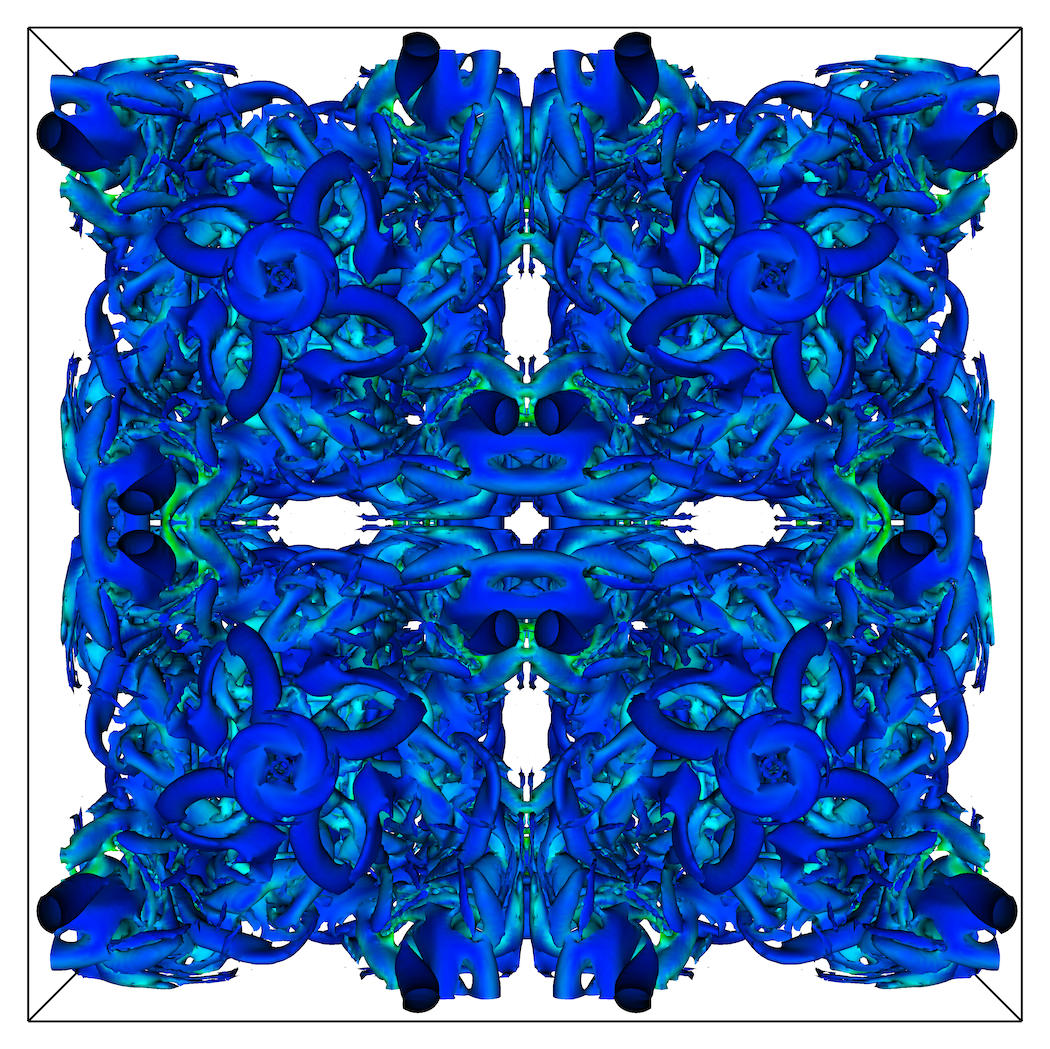}}
    \qquad
    \subfloat[]
    {\includegraphics[width=0.35\textwidth]{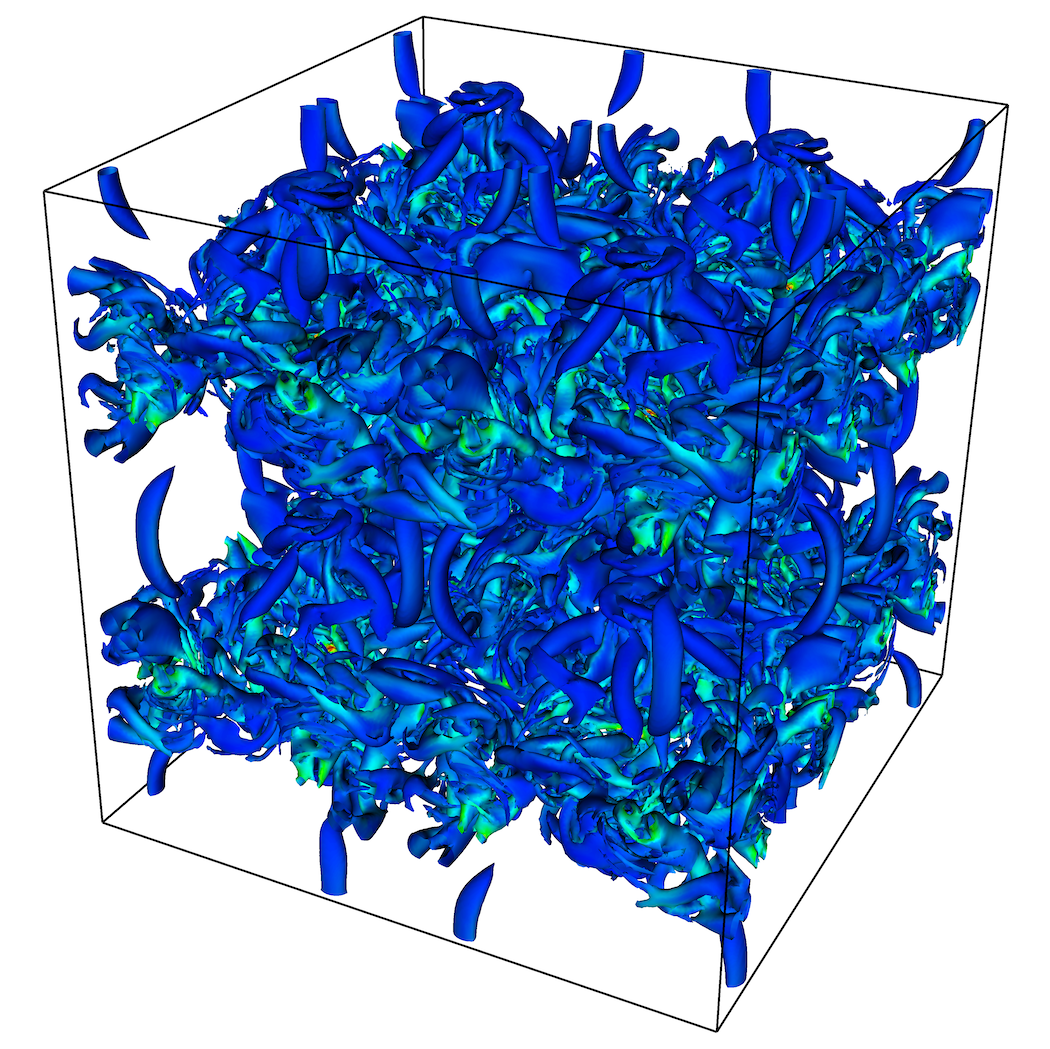}}
    \caption{Q-criterion iso-surface colored by enstrophy in the Taylor-Green vortex problem using $256^3$ points.}
    \label{fig:TGV_qcriterion}
\end{figure}

\subsection{Numerical schemes}
The problem is numerically computed on a 3D Cartesian uniform mesh using the staggered sixth order compact finite difference schemes and the sixth order compact interpolators \cite{lele1992compact, nagarajan2003robust}, as shown in the following two equations.
\begin{align}
    \frac{9}{62}f'_{i-1} + f'_i + \frac{9}{62}f'_{i+1}
    &= \frac{63}{62} \left(\frac{f_{i+1/2} - f_{i-1/2}}{ \Delta}\right)
    +  \frac{17}{62} \left(\frac{f_{i+3/2} - f_{i-3/2}}{3\Delta}\right)
    \\
    \frac{3}{10}f^I_{i-1} + f^I_i + \frac{3}{10}f^I_{i+1}
    &= \frac{3}{2}  \left(\frac{f_{i+1/2} + f_{i-1/2}}{2}\right)
    +  \frac{1}{10} \left(\frac{f_{i+3/2} + f_{i-3/2}}{2}\right)
\end{align}
where $f$, $f'$, and $f^I$ represent the original field, first derivative, and interpolated field, respectively; the subscript indicate the grid index in the corresponding direction; and $\Delta$ is the grid space in the corresponding direction. The primitive variables are all stored at the grid collocations, and all the fluxes in $\bm F$ and $\bm G$ are constructed at the staggered locations in the corresponding directions. The time advancement uses the standard fourth order Runge-Kutta method.

\bibliographystyle{elsarticle-num-names}
\bibliography{References}

\begin{thebibliography}{33}
\expandafter\ifx\csname natexlab\endcsname\relax\def\natexlab#1{#1}\fi
\providecommand{\url}[1]{\texttt{#1}}
\providecommand{\href}[2]{#2}
\providecommand{\path}[1]{#1}
\providecommand{\DOIprefix}{doi:}
\providecommand{\ArXivprefix}{arXiv:}
\providecommand{\URLprefix}{URL: }
\providecommand{\Pubmedprefix}{pmid:}
\providecommand{\doi}[1]{\href{http://dx.doi.org/#1}{\path{#1}}}
\providecommand{\Pubmed}[1]{\href{pmid:#1}{\path{#1}}}
\providecommand{\bibinfo}[2]{#2}
\ifx\xfnm\relax \def\xfnm[#1]{\unskip,\space#1}\fi
\bibitem[{Colonius and Lele(2004)}]{colonius2004computational}
\bibinfo{author}{T.~Colonius}, \bibinfo{author}{S.~K. Lele},
\newblock \bibinfo{title}{Computational aeroacoustics: progress on nonlinear
  problems of sound generation},
\newblock \bibinfo{journal}{Progress in Aerospace sciences}
  \bibinfo{volume}{40} (\bibinfo{year}{2004}) \bibinfo{pages}{345--416}.
\bibitem[{Lele(1992)}]{lele1992compact}
\bibinfo{author}{S.~K. Lele},
\newblock \bibinfo{title}{Compact finite difference schemes with spectral-like
  resolution},
\newblock \bibinfo{journal}{Journal of computational physics}
  \bibinfo{volume}{103} (\bibinfo{year}{1992}) \bibinfo{pages}{16--42}.
\bibitem[{Gottlieb and Orszag(1977)}]{gottlieb1977numerical}
\bibinfo{author}{D.~Gottlieb}, \bibinfo{author}{S.~A. Orszag},
  \bibinfo{title}{Numerical analysis of spectral methods: theory and
  applications}, \bibinfo{publisher}{SIAM}, \bibinfo{year}{1977}.
\bibitem[{Laizet and Lamballais(2009)}]{laizet2009high}
\bibinfo{author}{S.~Laizet}, \bibinfo{author}{E.~Lamballais},
\newblock \bibinfo{title}{High-order compact schemes for incompressible flows:
  A simple and efficient method with quasi-spectral accuracy},
\newblock \bibinfo{journal}{Journal of Computational Physics}
  \bibinfo{volume}{228} (\bibinfo{year}{2009}) \bibinfo{pages}{5989--6015}.
\bibitem[{Simens et~al.(2009)Simens, Jim{\'e}nez, Hoyas, and
  Mizuno}]{simens2009high}
\bibinfo{author}{M.~P. Simens}, \bibinfo{author}{J.~Jim{\'e}nez},
  \bibinfo{author}{S.~Hoyas}, \bibinfo{author}{Y.~Mizuno},
\newblock \bibinfo{title}{A high-resolution code for turbulent boundary
  layers},
\newblock \bibinfo{journal}{Journal of Computational Physics}
  \bibinfo{volume}{228} (\bibinfo{year}{2009}) \bibinfo{pages}{4218--4231}.
\bibitem[{Ghate and Lele(2017)}]{ghate2017subfilter}
\bibinfo{author}{A.~S. Ghate}, \bibinfo{author}{S.~K. Lele},
\newblock \bibinfo{title}{Subfilter-scale enrichment of planetary boundary
  layer large eddy simulation using discrete fourier-gabor modes},
\newblock \bibinfo{journal}{Journal of Fluid Mechanics} \bibinfo{volume}{819}
  (\bibinfo{year}{2017}) \bibinfo{pages}{494}.
\bibitem[{Uzun and Malik(2018)}]{uzun2018large}
\bibinfo{author}{A.~Uzun}, \bibinfo{author}{M.~R. Malik},
\newblock \bibinfo{title}{Large-eddy simulation of flow over a wall-mounted
  hump with separation and reattachment},
\newblock \bibinfo{journal}{AIAA Journal} \bibinfo{volume}{56}
  (\bibinfo{year}{2018}) \bibinfo{pages}{715--730}.
\bibitem[{Tritschler et~al.(2014)Tritschler, Olson, Lele, Hickel, Hu, and
  Adams}]{tritschler2014richtmyer}
\bibinfo{author}{V.~Tritschler}, \bibinfo{author}{B.~Olson},
  \bibinfo{author}{S.~Lele}, \bibinfo{author}{S.~Hickel},
  \bibinfo{author}{X.~Hu}, \bibinfo{author}{N.~A. Adams},
\newblock \bibinfo{title}{On the richtmyer--meshkov instability evolving from a
  deterministic multimode planar interface},
\newblock \bibinfo{journal}{Journal of Fluid Mechanics} \bibinfo{volume}{755}
  (\bibinfo{year}{2014}) \bibinfo{pages}{429--462}.
\bibitem[{Ryu and Livescu(2014)}]{ryu2014turbulence}
\bibinfo{author}{J.~Ryu}, \bibinfo{author}{D.~Livescu},
\newblock \bibinfo{title}{Turbulence structure behind the shock in canonical
  shock--vortical turbulence interaction},
\newblock \bibinfo{journal}{Journal of Fluid Mechanics} \bibinfo{volume}{756}
  (\bibinfo{year}{2014}).
\bibitem[{Jagannathan and Donzis(2016)}]{jagannathan2016reynolds}
\bibinfo{author}{S.~Jagannathan}, \bibinfo{author}{D.~A. Donzis},
\newblock \bibinfo{title}{Reynolds and mach number scaling in
  solenoidally-forced compressible turbulence using high-resolution direct
  numerical simulations},
\newblock \bibinfo{journal}{Journal of Fluid Mechanics} \bibinfo{volume}{789}
  (\bibinfo{year}{2016}) \bibinfo{pages}{669--707}.
\bibitem[{Olson et~al.(2011)Olson, Larsson, Lele, and
  Cook}]{olson2011nonlinear}
\bibinfo{author}{B.~J. Olson}, \bibinfo{author}{J.~Larsson},
  \bibinfo{author}{S.~K. Lele}, \bibinfo{author}{A.~W. Cook},
\newblock \bibinfo{title}{Nonlinear effects in the combined
  rayleigh-taylor/kelvin-helmholtz instability},
\newblock \bibinfo{journal}{Physics of Fluids} \bibinfo{volume}{23}
  (\bibinfo{year}{2011}) \bibinfo{pages}{114107}.
\bibitem[{Bodony and Lele(2005)}]{bodony2005using}
\bibinfo{author}{D.~J. Bodony}, \bibinfo{author}{S.~K. Lele},
\newblock \bibinfo{title}{On using large-eddy simulation for the prediction of
  noise from cold and heated turbulent jets},
\newblock \bibinfo{journal}{Physics of Fluids} \bibinfo{volume}{17}
  (\bibinfo{year}{2005}) \bibinfo{pages}{085103}.
\bibitem[{Wolf et~al.(2012)Wolf, Azevedo, and Lele}]{wolf2012convective}
\bibinfo{author}{W.~R. Wolf}, \bibinfo{author}{J.~L.~F. Azevedo},
  \bibinfo{author}{S.~K. Lele},
\newblock \bibinfo{title}{Convective effects and the role of quadrupole sources
  for aerofoil aeroacoustics},
\newblock \bibinfo{journal}{Journal of Fluid Mechanics} \bibinfo{volume}{708}
  (\bibinfo{year}{2012}) \bibinfo{pages}{502}.
\bibitem[{Ghaisas et~al.(2018)Ghaisas, Subramaniam, and
  Lele}]{ghaisas2018unified}
\bibinfo{author}{N.~S. Ghaisas}, \bibinfo{author}{A.~Subramaniam},
  \bibinfo{author}{S.~K. Lele},
\newblock \bibinfo{title}{A unified high-order eulerian method for continuum
  simulations of fluid flow and of elastic--plastic deformations in solids},
\newblock \bibinfo{journal}{Journal of Computational Physics}
  \bibinfo{volume}{371} (\bibinfo{year}{2018}) \bibinfo{pages}{452--482}.
\bibitem[{Shang(1999)}]{shang1999high}
\bibinfo{author}{J.~Shang},
\newblock \bibinfo{title}{High-order compact-difference schemes for
  time-dependent maxwell equations},
\newblock \bibinfo{journal}{Journal of Computational Physics}
  \bibinfo{volume}{153} (\bibinfo{year}{1999}) \bibinfo{pages}{312--333}.
\bibitem[{Nagarajan et~al.(2003)Nagarajan, Lele, and
  Ferziger}]{nagarajan2003robust}
\bibinfo{author}{S.~Nagarajan}, \bibinfo{author}{S.~K. Lele},
  \bibinfo{author}{J.~H. Ferziger},
\newblock \bibinfo{title}{A robust high-order compact method for large eddy
  simulation},
\newblock \bibinfo{journal}{Journal of Computational Physics}
  \bibinfo{volume}{191} (\bibinfo{year}{2003}) \bibinfo{pages}{392--419}.
\bibitem[{Wong and Lele(2017)}]{wong2017high}
\bibinfo{author}{M.~L. Wong}, \bibinfo{author}{S.~K. Lele},
\newblock \bibinfo{title}{High-order localized dissipation weighted compact
  nonlinear scheme for shock-and interface-capturing in compressible flows},
\newblock \bibinfo{journal}{Journal of Computational Physics}
  \bibinfo{volume}{339} (\bibinfo{year}{2017}) \bibinfo{pages}{179--209}.
\bibitem[{Subramaniam et~al.(2019)Subramaniam, Wong, and
  Lele}]{subramaniam2019high}
\bibinfo{author}{A.~Subramaniam}, \bibinfo{author}{M.~L. Wong},
  \bibinfo{author}{S.~K. Lele},
\newblock \bibinfo{title}{A high-order weighted compact high resolution scheme
  with boundary closures for compressible turbulent flows with shocks},
\newblock \bibinfo{journal}{Journal of Computational Physics}
  \bibinfo{volume}{397} (\bibinfo{year}{2019}) \bibinfo{pages}{108822}.
\bibitem[{Gander and Golub(1997)}]{gander1997cyclic}
\bibinfo{author}{W.~Gander}, \bibinfo{author}{G.~H. Golub},
\newblock \bibinfo{title}{Cyclic reduction—history and applications},
\newblock \bibinfo{journal}{Scientific computing (Hong Kong, 1997)}
  (\bibinfo{year}{1997}) \bibinfo{pages}{73--85}.
\bibitem[{Hockney(1965)}]{hockney1965fast}
\bibinfo{author}{R.~W. Hockney},
\newblock \bibinfo{title}{A fast direct solution of poisson's equation using
  fourier analysis},
\newblock \bibinfo{journal}{Journal of the ACM (JACM)} \bibinfo{volume}{12}
  (\bibinfo{year}{1965}) \bibinfo{pages}{95--113}.
\bibitem[{Buzbee et~al.(1970)Buzbee, Golub, and Nielson}]{buzbee1970direct}
\bibinfo{author}{B.~L. Buzbee}, \bibinfo{author}{G.~H. Golub},
  \bibinfo{author}{C.~W. Nielson},
\newblock \bibinfo{title}{On direct methods for solving poisson’s equations},
\newblock \bibinfo{journal}{SIAM Journal on Numerical analysis}
  \bibinfo{volume}{7} (\bibinfo{year}{1970}) \bibinfo{pages}{627--656}.
\bibitem[{Buneman(1969)}]{buneman1969compact}
\bibinfo{author}{O.~Buneman},
\newblock \bibinfo{title}{A compact non-iterative poisson solver},
\newblock \bibinfo{journal}{SUIPR report} \bibinfo{volume}{294}
  (\bibinfo{year}{1969}).
\bibitem[{Sweet(1974)}]{sweet1974generalized}
\bibinfo{author}{R.~A. Sweet},
\newblock \bibinfo{title}{A generalized cyclic reduction algorithm},
\newblock \bibinfo{journal}{SIAM Journal on Numerical Analysis}
  \bibinfo{volume}{11} (\bibinfo{year}{1974}) \bibinfo{pages}{506--520}.
\bibitem[{Sweet(1977)}]{sweet1977cyclic}
\bibinfo{author}{R.~A. Sweet},
\newblock \bibinfo{title}{A cyclic reduction algorithm for solving block
  tridiagonal systems of arbitrary dimension},
\newblock \bibinfo{journal}{SIAM Journal on Numerical Analysis}
  \bibinfo{volume}{14} (\bibinfo{year}{1977}) \bibinfo{pages}{706--720}.
\bibitem[{Swarztrauber(1974)}]{swarztrauber1974direct}
\bibinfo{author}{P.~N. Swarztrauber},
\newblock \bibinfo{title}{A direct method for the discrete solution of
  separable elliptic equations},
\newblock \bibinfo{journal}{SIAM Journal on Numerical Analysis}
  \bibinfo{volume}{11} (\bibinfo{year}{1974}) \bibinfo{pages}{1136--1150}.
\bibitem[{Hockney and Jesshope(1981)}]{hockney1981parallel}
\bibinfo{author}{R.~Hockney}, \bibinfo{author}{C.~Jesshope},
\newblock \bibinfo{title}{Parallel computers: Architecture},
\newblock \bibinfo{journal}{Programming and Algorithms, Adam Hilger, Bristol}
  (\bibinfo{year}{1981}).
\bibitem[{Zhang et~al.(2010)Zhang, Cohen, and Owens}]{zhang2010fast}
\bibinfo{author}{Y.~Zhang}, \bibinfo{author}{J.~Cohen}, \bibinfo{author}{J.~D.
  Owens},
\newblock \bibinfo{title}{Fast tridiagonal solvers on the gpu},
\newblock \bibinfo{journal}{ACM Sigplan Notices} \bibinfo{volume}{45}
  (\bibinfo{year}{2010}) \bibinfo{pages}{127--136}.
\bibitem[{Hirshman et~al.(2010)Hirshman, Perumalla, Lynch, and
  Sanchez}]{hirshman2010bcyclic}
\bibinfo{author}{S.~P. Hirshman}, \bibinfo{author}{K.~S. Perumalla},
  \bibinfo{author}{V.~E. Lynch}, \bibinfo{author}{R.~Sanchez},
\newblock \bibinfo{title}{Bcyclic: A parallel block tridiagonal matrix cyclic
  solver},
\newblock \bibinfo{journal}{Journal of Computational Physics}
  \bibinfo{volume}{229} (\bibinfo{year}{2010}) \bibinfo{pages}{6392--6404}.
\bibitem[{Seal et~al.(2013)Seal, Perumalla, and Hirshman}]{seal2013revisiting}
\bibinfo{author}{S.~K. Seal}, \bibinfo{author}{K.~S. Perumalla},
  \bibinfo{author}{S.~P. Hirshman},
\newblock \bibinfo{title}{Revisiting parallel cyclic reduction and parallel
  prefix-based algorithms for block tridiagonal systems of equations},
\newblock \bibinfo{journal}{Journal of Parallel and Distributed Computing}
  \bibinfo{volume}{73} (\bibinfo{year}{2013}) \bibinfo{pages}{273--280}.
\bibitem[{Subramaniam(2018)}]{subramaniam2018simulations}
\bibinfo{author}{A.~Subramaniam}, \bibinfo{title}{Simulations of shock induced
  interfacial instabilities including materials with strength},
  \bibinfo{publisher}{Stanford University}, \bibinfo{year}{2018}.
\bibitem[{Vazhkudai et~al.(2018)Vazhkudai, de~Supinski, Bland, Geist, Sexton,
  Kahle, Zimmer, Atchley, Oral, Maxwell et~al.}]{summit_2018}
\bibinfo{author}{S.~S. Vazhkudai}, \bibinfo{author}{B.~R. de~Supinski},
  \bibinfo{author}{A.~S. Bland}, \bibinfo{author}{A.~Geist},
  \bibinfo{author}{J.~Sexton}, \bibinfo{author}{J.~Kahle},
  \bibinfo{author}{C.~J. Zimmer}, \bibinfo{author}{S.~Atchley},
  \bibinfo{author}{S.~Oral}, \bibinfo{author}{D.~E. Maxwell}, et~al.,
\newblock \bibinfo{title}{The design, deployment, and evaluation of the coral
  pre-exascale systems},
\newblock in: \bibinfo{booktitle}{SC18: International Conference for High
  Performance Computing, Networking, Storage and Analysis},
  \bibinfo{organization}{IEEE}, \bibinfo{year}{2018}, pp.
  \bibinfo{pages}{661--672}.
\bibitem[{Bull and Jameson(2015)}]{bull2015simulation}
\bibinfo{author}{J.~R. Bull}, \bibinfo{author}{A.~Jameson},
\newblock \bibinfo{title}{Simulation of the taylor--green vortex using
  high-order flux reconstruction schemes},
\newblock \bibinfo{journal}{AIAA Journal} \bibinfo{volume}{53}
  (\bibinfo{year}{2015}) \bibinfo{pages}{2750--2761}.
\bibitem[{Towns et~al.(2014)Towns, Cockerill, Dahan, Foster, Gaither, Grimshaw,
  Hazlewood, Lathrop, Lifka, Peterson, Roskies, Scott, and
  Wilkins-Diehr}]{xsede}
\bibinfo{author}{J.~Towns}, \bibinfo{author}{T.~Cockerill},
  \bibinfo{author}{M.~Dahan}, \bibinfo{author}{I.~Foster},
  \bibinfo{author}{K.~Gaither}, \bibinfo{author}{A.~Grimshaw},
  \bibinfo{author}{V.~Hazlewood}, \bibinfo{author}{S.~Lathrop},
  \bibinfo{author}{D.~Lifka}, \bibinfo{author}{G.~D. Peterson},
  \bibinfo{author}{R.~Roskies}, \bibinfo{author}{J.~R. Scott},
  \bibinfo{author}{N.~Wilkins-Diehr},
\newblock \bibinfo{title}{{X}{S}{E}{D}{E}: Accelerating scientific discovery},
\newblock \bibinfo{journal}{Computing in Science \& Engineering}
  \bibinfo{volume}{16} (\bibinfo{year}{2014}) \bibinfo{pages}{62--74}.
  \URLprefix \url{doi.ieeecomputersociety.org/10.1109/MCSE.2014.80}.
  \DOIprefix\doi{10.1109/MCSE.2014.80}.

\end{thebibliography}
\end{document}